\documentclass[10pt,journal,compsoc]{IEEEtran}
\usepackage{amsmath}
\usepackage{amssymb}
\usepackage{amsthm}
\usepackage{multirow}
\usepackage{arydshln}
\usepackage{subcaption}

\ifCLASSOPTIONcompsoc
  \usepackage[nocompress]{cite}
\else
  \usepackage{cite}
\fi

\newcount\DraftStatus  
\usepackage{color}
\usepackage[nameinlink,noabbrev,capitalize]{cleveref}
\usepackage{bbm}

\ifnum\DraftStatus=1
\usepackage[draft,colorinlistoftodos,color=orange!30]{todonotes}
\else
\usepackage[disable,colorinlistoftodos,color=blue!30]{todonotes}
\fi 

\makeatletter

\newcommand{\Rmnum}[1]{\expandafter\@slowromancap\romannumeral #1@}
\makeatother

\newtheorem{definition}{Definition}
\usepackage{changes}
	\definechangesauthor[name={xm}, color=orange]{xm}
	\definechangesauthor[name={ql}, color=red]{ql}

\ifCLASSINFOpdf
\else
\fi

\begin{document}

\title{Causal Disentanglement for Semantics-Aware Intent Learning in Recommendation}

\author{\IEEEauthorblockN{Xiangmeng Wang\IEEEauthorrefmark{1},
Qian Li\IEEEauthorrefmark{1} \IEEEauthorrefmark{2},
Dianer Yu, 
Peng Cui, \IEEEmembership{Member, IEEE},
Zhichao Wang,
Guandong Xu\IEEEauthorrefmark{2}, \IEEEmembership{Member, IEEE} }

\IEEEcompsocitemizethanks{
\IEEEcompsocthanksitem X. Wang, D. Yu and G. Xu are with Data Science and Machine Intelligence Lab, Faculty of Engineering and Information Technology, University of Technology Sydney, New South Wales, Australia.
E-mail: {\{Guandong.Xu\}}@uts.edu.au
\protect\\
\IEEEcompsocthanksitem
Q. Li is with the School of Electrical Engineering, Computing and Mathematical
Sciences, Curtin University, Perth, Australia. E-mail: qli@curtin.edu.au.
\protect\\
\IEEEcompsocthanksitem
Z. Wang is with School of Electrical Engineering and Telecommunications, University of New South Wales, Sydney, Australia
\protect\\
\IEEEcompsocthanksitem P. Cui is with 
the Department of Computer Science and Technology in
Tsinghua University, Beijing 100084, China.\\
}
\thanks{* Both authors contributed equally to this research.}
\thanks{\dag Corresponding author.}
\thanks{This work is partially supported by the Australian Research Council (ARC) under Grant No. DP200101374, LP170100891, DP220103717 and LE220100078.}}

\markboth{Journal of \LaTeX\ Class Files,~Vol.~14, No.~8, August~2015}%
{X. Wang \MakeLowercase{\textit{et al.}}: Causal Disentanglement for Semantics-Aware Intent Learning}

\IEEEtitleabstractindextext{%
\begin{abstract}
Traditional recommendation models trained on observational interaction data have generated large impacts in a wide range of applications, it faces bias problems that cover users' true intent and thus deteriorate the recommendation effectiveness.
Existing methods tracks this problem as eliminating bias for the robust recommendation, e.g., by re-weighting training samples or learning disentangled representation.
The disentangled representation methods as the state-of-the-art eliminate bias through revealing cause-effect of the bias generation. However, how to design the semantics-aware and unbiased representation for users true intents is largely unexplored.
To bridge the gap, we are the first to propose an unbiased and semantics-aware disentanglement learning called \textbf{CaDSI}
(\textbf{Ca}usal \textbf{D}isentanglement for \textbf{S}emantics-Aware \textbf{I}ntent Learning) from a causal perspective.
Particularly, CaDSI explicitly models the causal relations underlying recommendation task, and thus produces semantics-aware representations via disentangling users true intents aware of specific item context.  
Moreover, the causal intervention mechanism is designed to eliminate confounding bias stemmed from context information, which further to align the semantics-aware representation with users true intent.
Extensive experiments and case studies both validate the robustness and interpretability of our proposed model.
\end{abstract}

\begin{IEEEkeywords}
Causal Disentanglement Learning, Semantics-aware Representation, Causal Intervention.
\end{IEEEkeywords}}

\maketitle

\IEEEdisplaynontitleabstractindextext

%
\IEEEpeerreviewmaketitle

\section{Introduction}\label{intro}
Recommender system (RS) has become a panacea for any scenario requiring personalized recommendations, to help users discover users' interested products from overwhelming alternatives.
Early works mainly adopt collaborative filtering (CF) methods~\cite{koren2009matrix,salakhutdinov2007restricted } to model user preference on items, based on historical user-item interactions (e.g., ratings, clicks). However, such user-item interaction usually exhibits bias that is entangled with users’ real interests, ignoring degrading the recommendation performance ultimately.
For instance, in movie recommendation, users are more likely to watch movies that
are watched by many people, which however is due to users’ conformity to other people, rather than stemming from users' real interests~\cite{zheng2021disentangling,wang2021deconfounded}.
Therefore, it is essential to capture users’ pure interests that are independent of the bias and thus can be leveraged to build high-quality recommender models.

Most existing works on bias-aware recommendation can be attributed into two categories. The first category adopts a re-weighting strategies on the  observed interaction samples, with the aim of imitating the scenario that samples are evenly distributed without bias~\cite{gruson2019offline,schnabel2016recommendations,wang2018deconfounded,ma2019learning}. 
One major limitation of these methods is that they merely
mitigate bias at the data level, but fail to answer the fundamental question: what are the root causes for bias amplification.
Another category of approaches that aim to disentangle user true-interest via inspecting cause-effect of the bias generation for the robust recommendation,
have recently gained much attention~\cite{wang2021deconfounded,li2021causal,zhang2021causal}.
Particularly, these works usually design a specific causal graph attributing the bias to a confounder. 
For example, social network is in fact a confounder for exposure bias, since it influences both users’ choice of movie watching
and their ratings~\cite{li2021causal}.
Apparently, the confounder introduces pseudo-intents, the ignorance of which definitely misguides the learning of user's true intent.
A widely used solution of these works is to learn the user representation that is forced to be independent of the confounder, with the aim of uncovering the user's true intents for final downstream recommendations.
Particularly, these works design a regularizer for the independence constraint via statistical
measures like $L_1$-inv, $L_2$-inv~\cite{10.1145/3442381.3449788} and distance correlation~\cite{ma2019learning}.
By explicitly disentangling the cause from a confounder, the representation uncovering user true intent can be learned for final recommendations.

Although promising improvements have been observed, existing approaches on disentangling user intent from bias still suffer two limitations:
First, most of them merely focus on user-item relationships, however the interaction data faces sparse issue in practical~\cite{ying2018graph,he2020lightgcn,shi2016survey}, leading to the difficulty of learning effective user or item representations.
Moreover, existing disentangled learning methods for user intents merely treat one user-item interaction record as an independent instance and neglect its rich context information.

We claim that the rich context information in the form of heterogeneous information can help to disentangle and interpret semantics-aware intents of users for the robust recommendation.  
For instance, higher-order graph structure like a meta path \emph{User-Movie-Actor-Movie-User} encodes the semantics interpretation of ``movies starring the same actor rated by the users''. In other words, without considering rich context information, disentangled learning fails in offering fine-grained interpretability in terms of item attributes for recommendation.

\begin{figure}[htbp]
\centering
\includegraphics[width=0.5\textwidth]{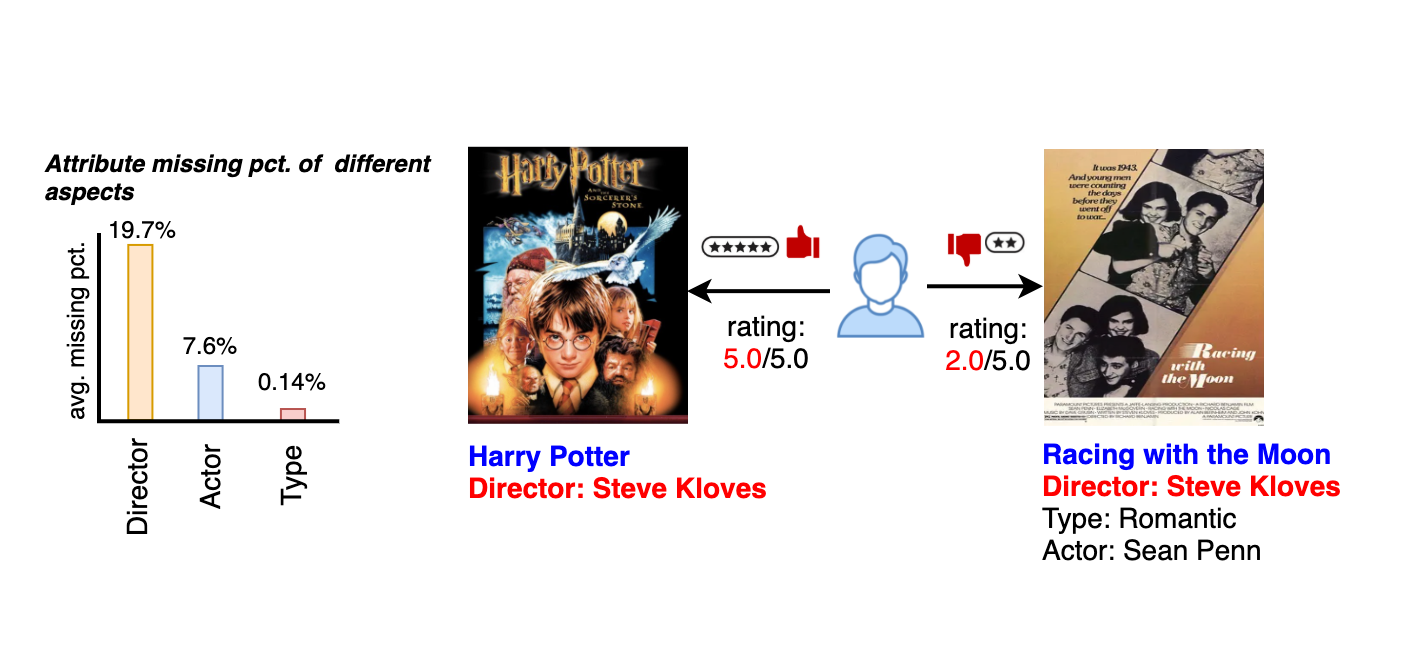}
\caption{An example of bias: movie \texttt{HP} contains only one aspect \emph{Director}, however, \emph{Actor} and \emph{Type} are missing.
The high rating of user on movie \texttt{HP} trains prediction model towards the user's preference on the director \emph{Steve Kloves}. 
In contrast, we observed that movie \texttt{RWM} with the same director received very low ratings from the same user.
}
\label{fig:exp1}
\end{figure}

Therefore, in this work, we propose to enhance user intent disentangled learning with heterogeneous information, which however is not trivial.
The heterogeneous information is complicated and consists of various types of data, e.g., item attributes. 
The complexity in heterogeneous information, e.g., the fact that items grouped by attributes (e.g., \emph{brand}) are frequently with skewed distributions, can bias the user preference and prediction score.
The skewed distribution is attributed to missing values of aspects, i.e., the number of non-missing aspects is not evenly distributed in observational dataset. 
An empirical study conducted on \texttt{Douban Movie} dataset can validate this claim by Figure~\ref{fig:exp1}: unobserved \emph{Director} aspect accounts for 19.7\% of items compared to \emph{Actor} accounting for 7.6\%.
That is, the skewed distribution of context aspect can easily bias the prediction model towards the majority group, even though their items have the same matching level (see example in Figure~\ref{fig:exp1}).

In this work, we attempt to tackle the challenge from a novel causal perspective, with the aim of developing a unbiased and interpretable disentangled approach on heterogeneous information, named \textbf{CaDSI} (\textbf{Ca}usal \textbf{D}isentanglement for \textbf{S}emantics-Aware \textbf{I}ntent Learning).
To make users' intents semantics-aware, we propose a pre-trained model as a first stage to leverage multiple item facets of heterogeneous information.
As a second stage, besides considering
the items directly interacted with the user, the higher-order interacted items via meta paths are exploited to disentangle user intents in a robust manner.
Finally, the pre-trained model together with disentangled learning is subsequently fine-tuned by the causal intervention. 
Thanks to the development of causal inference, the second module is designed to adopt causal intervention mechanism to eliminate the bias introduced by heterogeneous information.
With these two stages, our method can guide the unbiased and semantics-aware representations disentangling user intents for the robust recommendation.
Overall, the key contributions of this work are fourfold:

\begin{itemize}
    \item Fundamentally different from previous works, CaDSI is the first method that can disentangle the unbiased user's intents from a causal perspective, in the meanwhile endow each user intent with specific semantics under the disentanglement learning task.
    \item We design a novel causal graph for the qualitative analysis of causal relationships in recommendation, based on which a pre-trained model is designed.
    With heterogeneous information, the pre-trained model can semantically account for the item context influence towards the user intent.   
    \item To eliminate confounding bias stemmed from heterogeneous information, we perform the causal intervention on user representation and refine the pre-trained model for unbiased user intent learning.  
    \item We conduct extensive experiments to show that our CaDSI method outperforms state-of-the-art methods.
    The interpretability of our CaDSI is also validated by our empirical study. 
\end{itemize}

\section{Preliminary and Problem Formulation}\label{problem}
In this section, we will first present causal relations underlying the data generation mechanism of recommendation with access to the heterogeneous information.
Following this, we prove the existence of confounder in the heterogeneous information and discuss the consequences for ignoring the context bias bought by the confounder. 
We then introduce the causal intervention and discuss how intervention via back-door adjustment can control the context bias from a causal perspective. 
Finally, we introduce the important concepts and notations used in our approach and give the formal definition of our problem to be solved. 

\subsection{Problem definition}
We formulate our task as disentangling and interpreting users' intent based on Heterogeneous Information Network (HIN). The important concepts of HIN and the formal definition of our problem are given as follows. 

\begin{definition}
[Heterogeneous Information Network]
A Heterogeneous Information Network (HIN) is denoted as $\mathcal{G}=(\mathcal{V},\mathcal{E})$ with a node type mapping function: $\phi: \mathcal{V} \to \mathcal{A}$ and an edge type mapping function: $\psi: \mathcal{E} \to \mathcal{R}$, 
where $\mathcal{A}$ and $\mathcal{R}$ are the node type set and edge type set of $\mathcal{G}$, respectively. 
Each node $v \in \mathcal{V}$ and edge $e \in \mathcal{E}$ in a HIN belongs to one particular type in node/edge type sets $\mathcal{V}$/$\mathcal{E}$ with $\phi(v) \in \mathcal{A}$ and $\psi(e) \in \mathcal{R}$, where $|\mathcal{A}|+|\mathcal{R}| > 2 $.
\label{df:HIN}
\end{definition}

\begin{definition}
[Meta Path]\label{HIN} 
Meta path $\mathbf{p}$ is a path defined on the network schema
$T_{\mathcal{G}}=(\mathcal{A},\mathcal{R})$, and is denoted in the form of
$$
\mathbf{p} \triangleq (\mathcal{A}_1 \mathop \to\limits^{\mathcal{R}_1} \mathcal{A}_2 \mathop \to\limits^{\mathcal{R}_2}...\mathop \to\limits^{\mathcal{R}_l} \mathcal{A}_{o+1})
$$
which defines a composite  $\mathcal{R}=\mathcal{R}_1 \mathcal{R}_2 ... \mathcal{R}_o$ between type $\mathcal{A}_1$ and $\mathcal{A}_{o+1}$.
For simplicity, we use node type names to denote the meta path if no multiple relations exist between type pairs, as $\mathbf{p}=(\mathcal{A}_{1}\mathcal{A}_{2}...\mathcal{A}_{o+1})$.
Commonly, a HIN contains multiple meta paths, the meta path set is defined as $\mathcal{P}$ where each $\mathbf{p} \in \mathcal{P}$.
\end{definition}

Based on these important concepts, we collect the key notations in Table~\ref{tab:notion} and formulate the problem to be solved as follows.

\begin{definition}[Problem Definition]
Given user and item sets, we define an interaction matrix $\boldsymbol{y}\in \mathbb{R}^{m\times n}$
where entry $\boldsymbol{y}_{ui}=1$ indicates a user $u$ in user set interacts with an item $i$ in item set, otherwise $\boldsymbol{y}_{ui}=0$.
We also have additional contextual information about users and items, e.g., social relationships between users or item brands and categories, absorbing in $\mathcal{G}$ of Definition~\ref{df:HIN}.
Thus, we aim to learn the prediction function $P$ parameterized by $\Theta$,
such that $\hat{\boldsymbol{y}}_{ui}=P(u,i|\boldsymbol{y},\mathcal{G}; \Theta)$,
where $\hat{\boldsymbol{y}}_{ui}$
denotes the probability that user $u$ will engage with item $i$ conditional on the given $\boldsymbol{y}$ and $\mathcal{G}$.
\end{definition}

\subsection{A Causal View on Recommendation}

\begin{figure}[!htb]
 \includegraphics[width=1\linewidth]{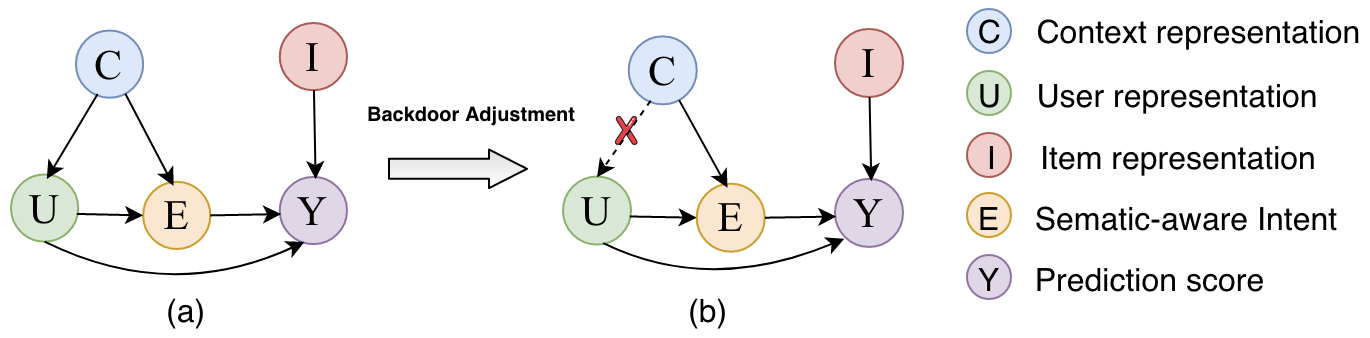}  
\centering
\caption{Causal disentanglement model for recommendation. We apply backdoor adjustment to remove the effect of confounder $C$ for $U$, as indicated by the red cross.}
\label{fig:SCM}
\end{figure}

\begin{figure*}[!htb]
\centering
 \includegraphics[width=0.95\linewidth]{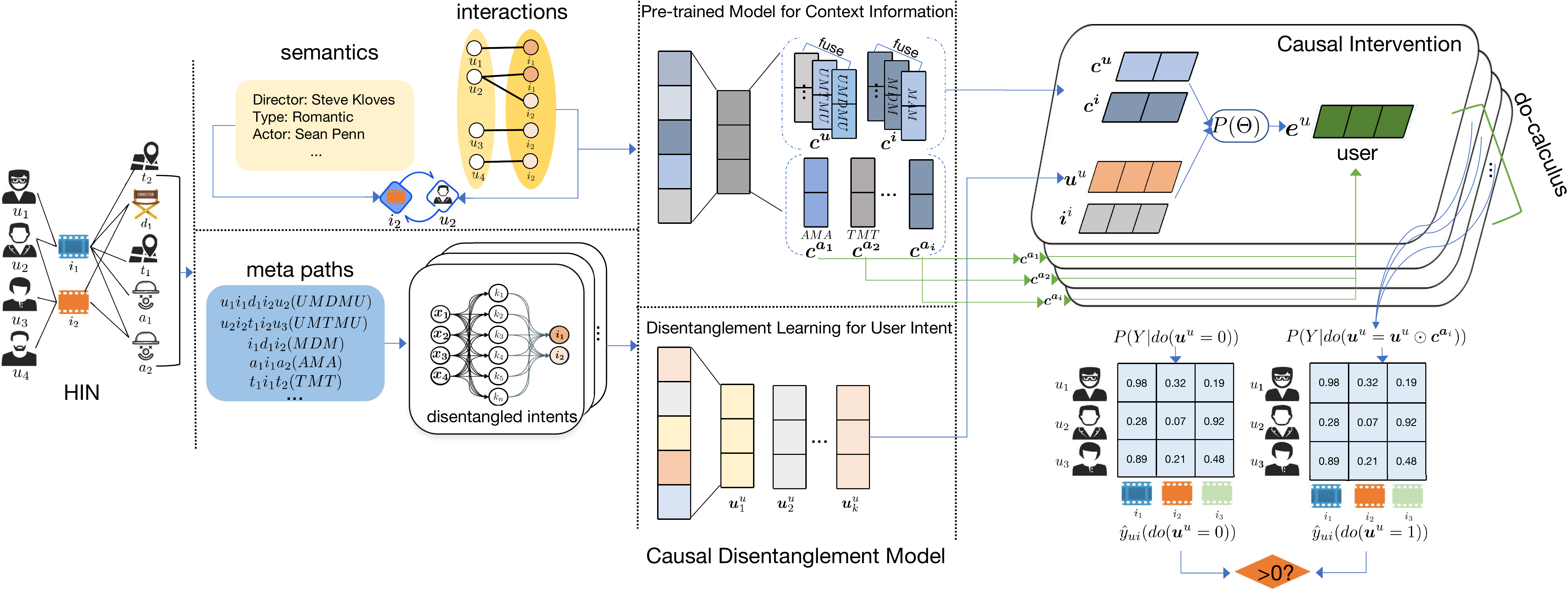}  
\caption{Overview of the proposed CaDSI. 
CaDSI takes HIN as the input, and passes the causal disentanglement model for learning intent-aware representations (\textit{cf.} Section~\ref{dis_task}); then use
the causal intervention (\textit{cf.} Section~\ref{fine-tuning}) for controlling the counfounding bias.}
\label{fig:frame}
\end{figure*}

\subsubsection{Structural Causal Model}
To illustrate the recommendation data generation mechanism seriously and soberly, we consider the structural causal model (SCM)~\cite{pearl2009causality} based on causality to reveal the true causal relations in recommendation. 
The basic idea of our proposed approach is to disentangle and interpret users' intent based on heterogeneous information. 
From a causal perspective, Figure~\ref{fig:SCM} (a) demonstrates the illustrative causal graph that offers an interpretable representation for disentangled recommender system, which consists of four variables including $\{U,C,E,Y\}$. In particular, as a directed acyclic graph, it can describe the generation mechanism of recommendation results and guide the design of recommendation methods. 
In the following, we explain the rationality of this causal graph at a higher-level.
\begin{itemize}

    \item $C$ as a confounder is the item aspects (e.g., movie genre) acquired from the heterogeneous information network. The representation of $C$ can be learned by a pre-trained model of context information, which retains semantics information of item aspects.

    \item $U$ denotes user representation which essentially reveals $k$ user intents.
    $U$ is presented in the form of $k$ chunked intent representation, where each chunk of representation reveals a piece of user intent, such as the user's special taste towards items' brand. 
   
    \item $I$ is item representations and each $I$ denotes the embedding of one item attribute (e.g. \emph{Genre}).
    \item $E$ is the semantics-aware intent representation generated by the context information from $C$ and the user representation $U$. 
    $E$ retains the information of 
    the user intent towards different item aspects. 
    \item $Y\in [0,1]$ is the recommendation probability for the user-item pair.
\end{itemize}

The directed edge represents the causal relation between two variables, in particular, the rationality of causal relations can be explained as follows.
\begin{itemize}
    \item $C \rightarrow U$: The prior knowledge $C$ of item aspects affect user representation $U$, which is reflected by the fact that users prefer the items 
    who have particular attributes (e.g., brand).

\item $(C, U)\rightarrow E$: Item context $C$ and user representation $U$ consist of the semantics-aware user intent representation.

\item $I\rightarrow Y$ : item representation by $I$ affects the recommendation probability $Y$.
\item $U\rightarrow Y$ : user's preference represented by $U$ affects the recommendation probability $Y$.

\item $U\rightarrow E\rightarrow Y$:  
the recommendation probability of item could be high if the user shows interest in the context of the item, e.g., the item type rather than the item.
For example, items whose type is "lipstick" are more likely to be purchased by the user $u$ whose gender is female.

\end{itemize}

\subsubsection{Adjusting Confounding Bias via Intervention}
From this causal graph, the semantic knowledge $C$ is a confounder between user representation $U$ and recommendation outcome $Y$, since $C$ is the common cause of $U$ and $Y$ by definitions in causal theory. 
The presence of confounder $C$ leads to the spurious correlation between $U$ and $Y$ if we ignore to account its causal effect into modeling, which is equal to the estimation of $P(Y \mid U)$.
In semantic knowledge-aware recommendation, the confounder $C$ (i.e., semantic knowledge) leads directly to the misguided recommendation probability $P(Y \mid U)$ that is biased towards items that have dominant item attributes. 
For example, as illustrated in Figure~\ref{fig:exp1}, we expect that the rating prediction of \texttt{RWM} is caused by both of the three item attributes, but not only the dominant item attribute \emph{director} which has a majority attribute popularity (i.e.,  the majority group).
In the language of causal inference, the conventional correlation $P(Y \mid U)$ fails to capture the true causality between $U$ and $Y$, 
because the prediction likelihood of $Y$ is conditional on not only $U$, but also the spurious correlation via (1) $C \rightarrow U\rightarrow Y$, i.e., prior knowledge $C$ determines the prediction likelihood through user representation $U$. For example, the undesirable and low-quality items in the specific attribute group will not attract users' intent, degrading recommendation accuracy. 
(2) $C \rightarrow E \rightarrow Y$.  i.e., the semantic-ware user intent representation $E$ derived from $C$ affects the prediction $Y$. 
Once users’ future interest in item attribute groups changes  (i.e., user interest drift), the recommendations will be biased.

To pursue the true causality between $U$ and $Y$, we should propose a causal intervention method $P(Y \mid do(U))$ to remove the confounding bias from $C$.
The $do(\cdot)$ operation~\cite{pearl2009causality} is to forcibly and externally assign a certain value to the variable $U$, which can be intuitively seen as removing the edge $C\rightarrow U$ and blocking the effect of $C$ on $U$ (as shown in Figure~\ref{fig:SCM} (b)). As the result, the prediction likelihood can be independent of its causes, so as to generate better recommendation performance that is free from the confounding bias.

\begin{table}[]
    \centering
    \caption{Key notations and descriptions.}
    \begin{tabular}{l|l}
    \hline
         Notation & Description \\\hline
         $\mathcal{G}$ & Heterogeneous Information Network (HIN)\\
         $\mathcal{V}$ & node set of HIN\\
         $\mathcal{A}$ & node type set of HIN\\
        $\mathcal{P}$ & meta paths set of HIN \\
        $\mathbf{p}$ & a meta path in $\mathcal{P}$\\
         $\boldsymbol{y}\in \mathbb{R}^{m\times n}$ & user item interaction matrix \\
         $\hat{\boldsymbol{y}}_{ui}$ & predicted interaction likelihood of user $u$ and item $i$\\
        $\boldsymbol{c}_{\boldsymbol{u}}$ & semantics-aware embedding for user $u$ \\
         $\boldsymbol{c}_{\boldsymbol{i}}$ & semantics-aware embedding for item $i$ \\
         $\boldsymbol{c}_{\boldsymbol{a}}$ & context information embedding for aspect $a$ \\
         $k$ & user intent number \\ 
        $\boldsymbol{l}$  & iteration number of graph disentangling module \\
         $L$ & $L$-th layer of graph disentangling module \\
         $\mathbf{S}_k (u,i) $ & intent score of $u$ and $i$ on intent $k$\\
         $\boldsymbol{u}^u$ & intent-aware embedding for $u$ \\
          $\boldsymbol{i}^i$ & embedding for item $i$ \\
        \hline
    \end{tabular}
    \label{tab:notion}
\end{table}

\section{Our method}
In this section, we first introduce motivation and the overall architecture
of the proposed model, which includes \emph{Causal Disentanglement Model} and \emph{Causal Intervention} as shown in Figure~\ref{fig:frame}. We then present the details of each component and
how they are applied to top-$N$ recommendation.

\subsection{Pre-trained Model for Learning Context Information}\label{learn_D}
The pre-trained model for learning context representation $C$ is a key component in the causal disentanglement model, which
aims to leverage side-information in the given HIN and construct expressive representations for users, items and aspects directly.
Specifically,
Given a HIN $\mathcal{G}=(\mathcal{V},\mathcal{E})$ and its corresponding meta paths set $\mathcal{P}$, we aim to learn semantics-aware representation (a.k.a., embedding) $\boldsymbol{c}_{\boldsymbol{u}} $ for each user node (i.e, \emph{User} type) that represents the semantic-aware embedding for user $u \in \mathcal{V}$, $\boldsymbol{c}_{\boldsymbol{i}}$ for each item node (i.e, \emph{Item} type) that represents the semantic-aware embedding for item $i \in \mathcal{V}$ and $\boldsymbol{c}_{\boldsymbol{a}}$ for each aspect node $a \in \mathcal{V}$ that represents context information representation of a specific type of aspect $a$ (e.g., \emph{Director}).

Towards this, a \emph{Heterogeneous Skip-Gram with Meta Path Based Random Walks} is designed to output a set of multinomial distributions, while each distribution corresponding to one type of node (i.e, \emph{User}, \emph{Item} and \emph{Aspect} type); 
the \emph{Meta Path Based Random Walks} is used to generate node sequences that capture the complex semantics reflected in a Heterogeneous Information Network (HIN), 
while \emph{Heterogeneous Skip-Gram} takes the generated node sequences as inputs and catches the heterogeneous neighborhood of a node for outputting the semantics-aware embeddings. Finally, the semantics-aware embeddings for users, items and aspects are given by aggregating every node representation under different meta paths by an \emph{Embedding Fusion} operation.

\subsubsection{Meta Path Based Random Walks} 
To generate node sequences that are able to capture both the semantics and structural correlations between different types of nodes. 
The \emph{Meta Path Based Random Walks}~\cite{dong2017metapath2vec}
is proposed to generate the node sequences traversed by random walkers over a HIN.
The basic idea is to put random walkers~\cite{grover2016node2vec} in a HIN to generate paths that constitute multiple types of nodes.
Specifically, given $\mathcal{G}=(\mathcal{V}, \mathcal{E}, \mathcal{A}, \mathcal{R}, \phi, \psi)$,
the node sequence $\mathbf{n}_\mathbf{p}=\{v_1,\cdots,v_{i+1}\}$ under a specific meta path $\mathbf{p}$ is generated according to the following distribution:

\begin{equation}\label{random_walk}
\begin{array}{l}
P\left(v_{i+1} \mid v_{i}, \mathbf{p} \right) \\
=\left\{\begin{array}{ll}
\frac{1}{\left|\mathcal{N}_{v_i}^{\left(\mathcal{A}_{o+1}\right)}\right|}, & (v_i, v_{i+1}) \in \mathcal{E} \text { and } \phi(v_{i+1})=\mathcal{A}_{o+1} \\
0, & \text { otherwise }
\end{array}\right.
\end{array}
\end{equation}
where $\mathcal{N}_{v_i}^{\left(\mathcal{A}_{o+1}\right)}$
is the first-order neighbor set for node $v_i$ whose type is $\mathcal{A}_{o+1}$; $v_{i+1}$ is the $i+1$-th node whose type is $\mathcal{A}_{o+1}$, and $v_i$ is the $i$-th node in the walk which belongs to type $\mathcal{A}_o$. By regulating $v_i \in \mathcal{A}_o$ while $v_{i+1} \in \mathcal{A}_{o+1}$, the node types sampled by random walkers is conditioned on the pre-defined meta path $\mathbf{p}$.

Following the pre-defined meta paths in Table~\ref{mp_scheme}, by performing the \emph{Meta Path Based Random Walks} on each meta path $\mathbf{p} \in \mathcal{P}$, 
we can obtain node sequences set for all meta paths as $\mathbf{n}^\mathcal{P}=\{\mathbf{n}_1,\cdots,\mathbf{n}_{|\mathcal{P}|}\}$.
As we care about sematic-aware embeddings for users, items and aspects, 
we then select meta paths $\mathbf{p}$ starting with \emph{user}, \emph{item} or \emph{Aspect} type and reorganize its corresponding node sequences $\mathbf{n}_\mathbf{p}$ into
\emph{user} type-specific set $\mathbf{n}(U)$ as $\mathbf{n}(U)=\{\mathbf{n}_1,\cdots,\mathbf{n}_m\}$,  \emph{item} type-specific set $\mathbf{n}(I)$ as $\mathbf{n}(I)=\{\mathbf{n}_1,\cdots,\mathbf{n}_n\}$ and aspect type-specific set $\mathbf{n}(A)=\{\mathbf{n}_1,\cdots,\mathbf{n}_h\}$.
We then use \emph{Heterogeneous Skip-Gram} to generate semantics-aware embeddings of node sequences in $\mathbf{n}(U)$, $\mathbf{n}(I)$ and $\mathbf{n}(A)$.

\subsubsection{Heterogeneous Skip-Gram}

Based on node sequences in set $\mathbf{n}(U)$, $\mathbf{n}(I)$ and $\mathbf{n}(A)$, we aim to leverage \emph{Heterogeneous Skip-Gram}~\cite{dong2017metapath2vec} to
learn node representations $\boldsymbol{c}_u$ and $\boldsymbol{c}_i$ and $\boldsymbol{c}_a$, which represent semantics-aware user, item and aspect representations of a specific node sequence $\mathbf{n}_i$ in $\mathbf{n}(U)$, $\mathbf{n}(I)$ and $\mathbf{n}(A)$, respectively. 
For concise purpose, we only present the learning process of the node representation $\boldsymbol{c}_u$, and analogously, we can obtain the representations of $\boldsymbol{c}_i$ and $\boldsymbol{c}_a$.

Specifically,
a \emph{Heterogeneous Skip-Gram} is designed to learn node representations by aggregating the heterogeneous neighborhood of the node in node sequence, while optimized through a node type-specific negative sampling~\cite{dong2017metapath2vec}.
Given each 
node sequence $\mathbf{n}_i$ in $\mathbf{n}(U)$ generated from Eq.~\eqref{random_walk}, 
the Skip-Gram model learns the semantics-aware embedding $\boldsymbol{c}_u$ of $\mathbf{n}_i$ by maximizing the probability of having the heterogeneous context $\mathcal{N}_{u}$ given a node $u$ as follows:

\begin{equation}\label{skip-gram}
\small
\mathcal{L}_{\theta}=
\sum_{u \in \mathcal{V}}  \sum_{u_c \in \mathcal{N}^{\mathcal{A}_i}_u} \sum_{\mathcal{A}_i \in \mathcal{A}} \left(\sigma\left({\boldsymbol{c}_{u}}^{T} \boldsymbol{c}_{u_c}\right) \prod_{w=1}^{W} \sigma\left({\boldsymbol{c}_{u}}^{T} \boldsymbol{c}_{w}\right) ; \theta \right)
\end{equation}
where $\mathcal{N}^{\mathcal{A}_i}_u$ denotes $u$’s neighborhood whose type is $\mathcal{A}_i$, $u_c$ is one node in the neighborhood set $\mathcal{N}_{u}$ of $u$,
$\boldsymbol{c}_u$ and $\boldsymbol{c}_{u_c}$ are latent vectors that correspond to the target node and context node representations of $u$ and $u_c$, and 
$\sigma(x)=1 / 1+\exp (-x)$.
$W$ is a parameter that determines the number of negative examples to be 
drawn per a positive example,
$\boldsymbol{c}_{w}$ is the sampled node's representation
within $W$ negative samples and $\theta$ is the model parameters of \emph{Heterogeneous Skip-Gram}.
Finally, $\boldsymbol{c}_u$  for each node sequence $\mathbf{n}_i$ in $\mathbf{n}(U)$ are estimated by applying gradient descent algorithm~\cite{bottou2012stochastic} with respect to the objective in Eq.~\eqref{skip-gram}.

\subsubsection{Embedding Fusion}
Since we have multiple node sequences in $\mathbf{n}(U)$, $\mathbf{n}(I)$ and $\mathbf{n}(A)$, while each learned representation $\boldsymbol{c}_u$, $\boldsymbol{c}_i$ and $\boldsymbol{c}_a$ is the semantic-aware embedding of each node sequence $\mathbf{n}_i$ in $\mathbf{n}(U)$, $\mathbf{n}(I)$ and $\mathbf{n}(A)$ respectively, we therefore perform embedding fusion
to aggregate every representations $\boldsymbol{c}_u$, $\boldsymbol{c}_i$ and $\boldsymbol{c}_a$ into an uniform manner categorized by their node types so as to guide the recommendation task. 
The reason why we fuse the individual embedding of 
each node sequence is quite straightforward. 
Firstly, in recommendation system, the optimization goal is to learn effective representations for users and items.
Hence, it requires a principled fusion way to transform node embeddings w.r.t. different meta paths relating \emph{user} type or \emph{item} type into a more suitable form for later recommendation tasks.
Secondly, the context information across meta paths starting with an aspect type should be further arranged into an uniform embedding space, representing one piece of semantics meaning, e.g., the aspect of the object been ``Director''.

The embedding fusion is implemented as a liner combination function defined as follows: 

\begin{equation}\label{learn_c}
 \begin{aligned}
\boldsymbol{c}_{\boldsymbol{u}} \leftarrow \frac{1}{\left|\boldsymbol{c}_{u}(U)\right|} \sum_{j=1}^{\left|\boldsymbol{c}_{u}(U)\right|}\left(\mathbf{M} \cdot \boldsymbol{c}_{u}^j+b\right) \\
\boldsymbol{c}_{\boldsymbol{i}} \leftarrow \frac{1}{\left|\boldsymbol{c}_{i}(I)\right|} \sum_{j=1}^{\left|\boldsymbol{c}_{i}(I)\right|}\left(\mathbf{M} \cdot \boldsymbol{c}_{i}^j+b\right) \\
\boldsymbol{c}_{\boldsymbol{a}} \leftarrow \frac{1}{\left|\boldsymbol{c}_{a}(A)\right|} \sum_{j=1}^{\left|\boldsymbol{c}_{a}(A)\right|}\left(\mathbf{M} \cdot \boldsymbol{c}_{a}^j+b\right)
\end{aligned}
\end{equation}
where $\boldsymbol{c}_{u}(U)$ is the user node representation set which absorb the node representations of user $u$
and $\boldsymbol{c}_u (U)=\{\boldsymbol{c}_{u}^1,\cdots,\boldsymbol{c}_{u}^{m}\}$.
Correspondingly, 
the item and aspect node representation sets $\boldsymbol{c}_i (I) =\{\boldsymbol{c}_{i}^1,\cdots,\boldsymbol{c}_{i}^{n}\}$ and $\boldsymbol{c}_a (A) =\{\boldsymbol{c}_{a}^1,\cdots,\boldsymbol{c}_{a}^{h}\}$ are established for item $i$ and aspect $a$.
$\mathbf{M}$ is a linear combination transformation matrix~\cite{weisberg2005applied} and $\boldsymbol{b}$ is the error term.
Through Eq.~\eqref{learn_c}, $\boldsymbol{c}_{\boldsymbol{u}}$, $\boldsymbol{c}_{\boldsymbol{i}}$ and $\boldsymbol{c}_{\boldsymbol{a}}$ 
can be learned as final semantics-aware representations for a user $u$ and an item $i$ and
context information representation for aspect $a$, respectively.

\subsection{Disentanglement Learning for User Intent}\label{learn_X}

Inspired by recent achievements on GNNs~\cite{berg2017graph,wang2019neural,ying2018graph,he2020lightgcn}, we propose a GNN-based disentangling module
to learn user representations that can essentially reveal $k$ user intents.
Specifically, 
the $L$-layer disentangling module exploits the high-order connectivities among user-item interaction graph and initializes intent-ware embeddings by separating each user/item embedding into $k$ chunks. 
Then, an interactive update rule that computes the importance scores of intent-aware user-item interactions is designed to refine intent-aware embeddings, so as to disentangle the holistic interaction graph into $k$ intent-ware sub-graphs. 
Thereafter, each intent-aware embedding chunk are stacked by embedding propagation in the current layer, serving as the holistic intent-aware embedding $\boldsymbol{u}^u$ and $\boldsymbol{i}^i$ for user $u$ and item $i$, where each intent-aware embedding $\boldsymbol{u}^u$ and $\boldsymbol{i}^i$ is composed of $k$ independent chunks:

    \begin{equation}\label{u_initit}
    \begin{split}
            \boldsymbol{u}^u=[\boldsymbol{u}_{1}^u,\cdots,\boldsymbol{u}_{k}^u],\quad \boldsymbol{i}^i=[\boldsymbol{i}_{1}^i,\cdots,\boldsymbol{i}_{k}^i]
    \end{split}
    \end{equation}
Each chunked representation $\boldsymbol{u}^{u}_{k}  \in \mathbb{R}^{\frac{d}{k}} $ and $\boldsymbol{i}^{i}_{k}  \in \mathbb{R}^{\frac{d}{k}} $is built upon the intent-aware interactions between user $u$ and its preferred items under intent $i$. 
We ultimately sum up the intent-aware representations at each intent $k$ of all $L$ layers,
the final layer outputs the $k$-th chunked intent-aware representations $\boldsymbol{u}^{u}_{k}$ and $\boldsymbol{i}^i_k$:

\begin{equation}\label{learn_u}
\boldsymbol{u}^{u}_{k}=\boldsymbol{u}^{u}_{k}{(1)}+\cdots+\boldsymbol{u}^{u}_{k}{(L)},\quad
\boldsymbol{i}^{i}_{k}=\boldsymbol{i}^{i}_{k}{(1)}+\cdots+\boldsymbol{i}^{i}_{k}{(L)}
\end{equation}

The detailed operations are given as follows.

\subsubsection{Initialization}

As ID embedding captures intrinsic characteristics of users, we separate the ID embeddings of user $u$ into $k$ chunks and associate each chunk with a latent
intent, serving as the initialization of the intent-aware embeddings $\boldsymbol{u}^u$ in Eq.~\eqref{u_initit}:
\begin{equation}\label{chunkedID}
\begin{aligned}
\boldsymbol{x}=\left[\boldsymbol{x}_{1}, \boldsymbol{x}_{2}, \cdots, \boldsymbol{x}_{k}\right] \\
\end{aligned}
\end{equation}

Thereafter, we initialize the importance score
$\mathbf{S}_{k}(u, i)$.
The $\mathbf{S}_{k}(u, i)$ is the importance score of the interaction between $u$ and $i$ with respect to the $k$-th intent. 
Such $\mathbf{S}_{k}(u, i)$ can be seen as the indicator of whether $u$ should interact with $i$ under a specific intent $k$.
Thus, by learn $\mathbf{S}_{k}(u, i)$ for aspect $k \in \{1,\cdots, k\}$, we can construct an intent-aware sub-graph at intent $k$. 
Such that, the embedding propagation can output intent-aware representation at intent $k$ for all users based on the corresponding intent-aware sub-graph. 
We uniformly initialize importance score $\mathbf{S}_{k}(u, i)$ as $\mathbf{S}_{k}(u, i)=\frac{1}{k}$ which presumes the equal contributions of intents at the start of modeling.

\subsubsection{Iterative Update Rule} 
An iterative update rule is then designed to update the importance score $\mathbf{S}_{k}(u, i)$ of user-item connections under aspect $k$ within $\boldsymbol{l}$ iterations, so as to disentangle interaction sub-graph at intent $k$ to refine each intent-aware embedding chunk $\boldsymbol{x}_k$ in Eq.~\eqref{chunkedID}.
Note that $\boldsymbol{x}_k \in \boldsymbol{x}$ in Eq.~\eqref{chunkedID} serves as the initialized representation chunk of $\boldsymbol{u}^u_k \in \boldsymbol{u}^u$ in Eq.~\eqref{u_initit} and is used to memorize the update value during iteration, the final $\boldsymbol{x}_k$ is assigned to each $\boldsymbol{u}^u_k$ as the final intent-aware representation chunk of user $u$.

In particular, we set $\boldsymbol{l}$
iterations in the interactive update.
At each iteration, for the target interaction $(u,i)$, we firstly
normalize the score vector $\mathbf{S}_{k}(u, i) \mid \forall k \in \{1, \cdots, k\}\}$ over all intents into $\tilde{\mathbf{S}}_{k}$ through a softmax function: 

\begin{equation}\label{intent_s}
\tilde{\mathbf{S}}_{k}^{\boldsymbol{l}}(u, i)=\frac{\exp \left( \mathbf{S}_{k}^{\boldsymbol{l}}(u, i)\right)}{\sum_{k^{\prime}=1}^{k} \exp \left(  \mathbf{S}_{k^{\prime}}^{\boldsymbol{l}}(u, i)\right)}
\end{equation}
which is capable of illustrating which intents should get more attention to explain each user behavior $(u,i)$.
We then perform embedding propagation over individual intent-aware graphs whose representation is denoted by $\boldsymbol{x}_{k}$, $\boldsymbol{i}_{k}^{i}$ and its adjacency matrix is denoted by $\tilde{\mathbf{S}}_{k}$ in Eq.~\eqref{intent_s},
such that the information of all individual intent-aware graphs are encoded into the learned representations.
The weighted sum aggregator is defined as:
\begin{equation}
\boldsymbol{x}_{k}^{\boldsymbol{l}}=\sum_{i \in \mathcal{N}_{u}} \frac{\tilde{\mathbf{S}}_{k}^{\boldsymbol{l}}(u, i)}{\sqrt{D_{k}^{\boldsymbol{l}}(u) \cdot D_{k}^{\boldsymbol{l}}(i)}} \cdot \boldsymbol{i}_{k}^{i}
\end{equation}
where $D_{k}^{\boldsymbol{l}}(u)=\sum_{i^{\prime} \in \mathcal{N}_{u}} \tilde{\mathbf{S}}_{k}^{\boldsymbol{l}}\left(u, i^{\prime}\right) $ and $D_{k}^{\boldsymbol{l}}(i)=\sum_{u^{\prime} \in \mathcal{N}_{i}} \tilde{\mathbf{S}}_{k}^{\boldsymbol{l}}\left(u^{\prime}, i\right)$ are
the degrees of user $u$ and item $i$, respectively.

Then we interactively update the intent-aware graphs.
Intuitively, historical items of users driven by the same intent tend to have the similar chunked representations, such goal can be achieved by encouraging users and items among the same intent to have stronger relationships.
We hence iteratively update the interaction importance score $\mathbf{S}_{k}^{\boldsymbol{l}}(u, i)$ 
in order to strengthen the degree
between the centroid $u$ and its neighbor $i$ under intent $k$, as follows:

\begin{equation}\label{one_layer}
\mathbf{S}_{k}^{\boldsymbol{l}+1}(u, i)=\mathbf{S}_{k}^{\boldsymbol{l}}(u, i)+\boldsymbol{x}_{k}^{\boldsymbol{l} \top} \tanh \left(\boldsymbol{i}_{k}^{i}\right)
\end{equation}
where $\boldsymbol{x}_{k}^{\boldsymbol{l} \top} \tanh \left(\boldsymbol{i}_{k}^{i}\right)$ considers the affinity between $\boldsymbol{x}_{k}^{\boldsymbol{l}}$ and $\boldsymbol{i}_{k}^{i}$,
and $tanh$~\cite{sharma2017activation} is a nonlinear activation function to increase the representation ability of model.

After $\boldsymbol{l}$ iterations,
$\boldsymbol{u}^{u}{(1)}=\boldsymbol{x}^{\boldsymbol{l}}{(1)}$
for user $u$ is obtained in the current layer, 
where each chuncked representation $\boldsymbol{u}^{u}_{k}{(1)}=\boldsymbol{x}_{k}^{\boldsymbol{l}}{(1)}$ denotes the chunked embedding of $\boldsymbol{u}^{u}{(1)}$ on the intent $k$ corresponds to the $k$-th dimension of Eq.~\eqref{u_initit}.

\subsubsection{Layer Combination}
To explore high-order connectivity between users and items,
we recursively formulate the representation of $L$-th layer as: 

\begin{equation}\label{layer}
\boldsymbol{u}^{u}_k{(L)}=g\left(\boldsymbol{x}^{\boldsymbol{l}}_{k}{(L-1)},\left\{\boldsymbol{i}^i_{k}{(L-1)} \mid i \in \mathcal{N}_{u}\right\}\right) 
\end{equation}
where $\boldsymbol{u}^{u}_{k}{(L)}$ and  $\boldsymbol{i}^{i}_{k}{(L)}$ 
are the representations of user $u$ and item
$i$ on the $k$-th intent at layer $L$, $\boldsymbol{x}^{\boldsymbol{l}}_{k}{(L-1)}$ is the chunked embedding of $k$-th intent at $L-1$-th layer of $\boldsymbol{u}^{u}_k{(L-1)}$. Note that
$g(\cdot)$ is a fully connection layer that memorizes the information propagated from the $(L-1)$-order
neighbors of $u$. 

Finally, 
after $L$ layers, we sum up intent-aware representations at
different layers as the final representation of the $k$-th chunk of Eq.~\eqref{u_initit}, as 
$\boldsymbol{u}^{u}_{k}=\boldsymbol{u}^{u}_{k}{(1)}+\cdots+\boldsymbol{u}^{u}_{k}{(L)}$,
where $\boldsymbol{u}^{u}_{k}$ donates the intent-aware representation for user $u$ at intent $k$.
Analogously, we can establish the final intent-aware embedding chunk $\boldsymbol{i}^{i}_{k}$ follow the definition in Eq.~\eqref{learn_u}.

\subsection{Semantics-aware Intent Learning}\label{dis_task}
 
Having obtained the intent-aware embeddings $\boldsymbol{u}^u$, $\boldsymbol{i}^i$ from \emph{disentanglement learning for User Intent},
to provide semantics information to the recommendation task, 
our method takes semantics factors $\boldsymbol{c}_{\boldsymbol{u}}$ and  $\boldsymbol{c}_{\boldsymbol{i}}$ from \emph{Pre-trained Model for Context Information} as the auxiliary input.
To facilitate the usage of semantics factors,
we design an operator to instantiate a semantics-aware intent representation $\boldsymbol{e}$, which denotes the user intent towards different aspects.
By learning $\boldsymbol{e}$, the effect of semantics factors towards user intent can be incorporated into the final updated user representation $\boldsymbol{u}^u$, such that $\boldsymbol{u}^u$ can be easily plugged into the backdoor adjustment to alleviate bias.
Specifically, a second-order Factorization Machine (FM)~\cite{5694074} module is used to instantiate $\boldsymbol{e}$: 

\begin{equation}\label{learn_e}
\boldsymbol{e}=\sum_{a=1}^{d} \sum_{b=1}^{d} \boldsymbol{u}^u_{a} \boldsymbol{c}_{\boldsymbol{i}_{a}} \odot \boldsymbol{c}_{\boldsymbol{u}_{b}} \boldsymbol{i}^i_{b}
\end{equation}
where $\odot$ denotes the element-wise product between each latent vector, such that the learned $\boldsymbol{e}$ captures the interactions between the intent-aware representation $\boldsymbol{u}^u$/$\boldsymbol{i}^i$ and semantics factors in $\boldsymbol{c}_{\boldsymbol{u}}$/$\boldsymbol{c}_{\boldsymbol{i}}$.

Next, the semantics-aware intent representation $\boldsymbol{e}$ can be incorporated into recommender models as one additional user representation.
Formally, we use the collaborative filtering to calculate the prediction score $\hat{\boldsymbol{y}}_{ui}$ given user and item ID representations, as follows:
\begin{equation}\label{calcu_y}
\hat{\boldsymbol{y}}_{ui}=f(\boldsymbol{u}, \boldsymbol{i},\boldsymbol{e}) = \delta \boldsymbol{u}^{\top} \boldsymbol{i}+(1-\delta) \boldsymbol{e}^{\top} \boldsymbol{i}
\end{equation}
where $\boldsymbol{u}$ and $\boldsymbol{i}$ are the ID embeddings given by id mapping techniques such as Multi-OneHot~\cite{zhang2019stylistic}, and $\delta$ is the coefficient that describes how much each component contributes to the prediction score.  
Then we use the pairwise BPR loss~\cite{lian2020personalized} to optimize the model parameters $\Theta$.
Specifically, BPR loss encourages the prediction of a user’s historical items to be higher than those of unobserved items:

\begin{equation}\label{BPR}
\mathcal{L}_{\mathrm{BPR}}=\sum_{(u, i, j) \in O}-\ln \sigma\left(\hat{\boldsymbol{y}}_{ui}-\hat{\boldsymbol{y}}_{uj}\right)+\lambda\|\Theta\|_{2}^{2}
\end{equation}
where $\boldsymbol{u}$, $\boldsymbol{i}$ and $\boldsymbol{j}$ are the ID embeddings of user $u$, item $i$ and item $j$, $O=\left\{(u, i, j) \mid(u, i) \in O^{+},(u, j) \in O^{-}\right\}$ denotes the training dataset involving the observed interactions $O^{+}$ and unobserved counterparts $O^{-}$; 
$\sigma(\cdot)$ is sigmoid function; 
$\lambda$ is the coefficient controlling regularization.

\subsection{Causal Intervention for Debiasing}\label{fine-tuning}

The context information as a confounder tends to introduce bad effect on both user representation and prediction score.
To make context information beneficial for semantics-aware intent learning, we resort to causal technique, backdoor adjustment\cite{pearl2009causality} to  adjust the representation mechanism in disentangled causal model.
By doing this, we aim to produce unbiased user representation and apply the modified one to recommendation task.

\subsubsection{Backdoor Adjustment}
Note that previous recommendation methods build a predictive model $P(Y \mid U)$ from the passively collected interaction dataset, which neglects the effect of confounder $C$, thus leading to a spurious correlation between users and items. The spurious correlation is harmful to most users because
the items in the majority group are likely to dominate the recommendation list and narrow down the user interests.

According to the theory of backdoor adjustment~\cite{pearl2009causality}, the target of our method is to remove the bad effect of context information $C$ on user representation $U$.
Instead of $P(Y \mid U)$, we formulate the predictive model as $P(Y|do(U=u))$ to account for the effect of confounder.
Based on the graph in Figure.~\ref{fig:SCM}, we first need to formulate the causal effect between variables by causal intervention, which is denoted as the $do(\cdot)$ operation~\cite{pearl2009causality}. $do(U=u)$ is to forcibly and externally assign a certain value to the variable $U$, which can be intuitively seen as removing the edge $C\rightarrow U$ and blocking the effect of $C$ on $U$, making its value independent of its causes (\textit{cf.} Figure.~\ref{fig:SCM}). By applying $do$ operation, we can estimate the effect of $C$ on $Y$ as 
\begin{equation}\label{do}
P(Y|do(U=u))-P(Y|do(U=0))
\end{equation}
where $P(Y|do(U=0))$ denotes the null intervention, e.g., the baseline compared to $U=u$.
In the physical world, $P(Y|do(U=u))$ corresponds to actively manipulating the aspects or attributes in the item.

Our implementation is inspired from two inherent properties of any Heterogeneous Network Embedding method (e.g., metapath2vec++~\cite{dong2017metapath2vec}).
First, by passing a meta-path into the pre-trained context model from Eq.~\eqref{learn_c}, we have the context embedding for an aspect,  denoted as $\boldsymbol{c}_{\boldsymbol{a}}$.
Aggregating context embeddings for all aspects into the aspect representation set $C$, we can obtain the unified aspect representation $C$, where each element of $C$, i.e., $\boldsymbol{c}_{\boldsymbol{a}}$, representing one semantic aspect (e.g., ``Director'') is computed by Eq.~\eqref{learn_c}. 
As $C$ is no longer correlated with $\boldsymbol{u}^u$ by removing the edge $C\rightarrow U$, the causal intervention makes $\boldsymbol{u}^u$ have a fair opportunity to incorporate every context $\boldsymbol{c}_{\boldsymbol{a}}$ into the prediction of $\hat{\boldsymbol{y}}_{ui}$, subject to a prior $P(C=\boldsymbol{c}_{\boldsymbol{a}})$.
Second, prevailing pre-trained models use a specific task (in our method is the semantics-aware intent learning in Section~\ref{dis_task}) as the objective, 
the representations trained from it can be considered as the
distilled information $\boldsymbol{u}^u$ that waits to adjust by $do(U=\boldsymbol{u}^{u} \odot C)$.

By far, we have the context information $C$ for all aspects,
where each $\boldsymbol{c}_{\boldsymbol{a}} \in C $ represents context embedding of one aspect.
We also have the refined intent representation $\boldsymbol{u}^u$ that waits to be adjusted from Eq.~\eqref{BPR} as $U$. 
Next, we will detail the proposed causal intervention by providing implementation for
Eq.~\eqref{do}.

The overall backdoor adjustment is achieved through:

\begin{equation}\label{backdoor}
\footnotesize
\begin{split}
   &P(Y \mid U, d o(U=u))-P(Y \mid d o(U=0))\\
&= \sum_{C} \left(P\left(Y \mid d o\left(U=\boldsymbol{u}^{u} \odot C \right)\right)-P\left(Y \mid d o\left(U=0\right)\right)\right) P\left(C=\boldsymbol{c}_{\boldsymbol{a}}\right) \\
&= \frac{1}{N} \sum_{i=1}^{N} \left(P\left(\hat{\boldsymbol{y}}_{ui} \mid \boldsymbol{u}^{u} \odot C \right) - P\left(\hat{\boldsymbol{y}}_{ui} \mid \boldsymbol{u}^{u} \right)\right)
\end{split}
\end{equation}
where each component in Eq.~\eqref{backdoor} is designed by:
\begin{itemize}
   
    \item $C=\boldsymbol{c}_{\boldsymbol{a}}$ indicates the confounder $C$ is set as one context embedding $\boldsymbol{c}_{\boldsymbol{a}}$.
    
    \item $P(Y|U,do(U=u))= P\left(\hat{\boldsymbol{y}}_{ui} \mid\boldsymbol{u}^{u} \odot C\right)$.
    The selected context embedding $C$ is concatenated with  $\boldsymbol{u}^u$ by the element-wise product $\odot$, which serves as the adjustment value for the prediction of $\hat{\boldsymbol{y}}_{ui}$.
    \item $P(C=\boldsymbol{c}_{\boldsymbol{a}})$ is the prior distribution of different item aspect, by defining $P(C=\boldsymbol{c}_{\boldsymbol{a}}) = 1/N$, we can assume a uniform prior for the adjusted features, where $N$ is the total number of aspect type. 
   
\end{itemize}

We then utilize an inference strategy to adaptively fuse the prediction scores from the $P\left(\hat{\boldsymbol{y}}_{ui} \mid \boldsymbol{u}^{u} \odot C\right)$ and $P\left(\hat{\boldsymbol{y}}_{ui} \mid \boldsymbol{u}^{u} \right)$.
Specifically, we first train the recommender model by $P\left(Y \mid d o\left(\hat{\boldsymbol{y}}_{ui}=\boldsymbol{u}^{u} \odot C \right)\right)$
and $P\left(Y \mid d o\left(\hat{\boldsymbol{y}}_{ui}=0\right)\right)$ and obtain $\hat{\boldsymbol{y}}_{C}$ and $\hat{\boldsymbol{y}}$, respectively.
Then, the adjusted prediction scores $\hat{\boldsymbol{y}}_{C}$ and unadjusted prediction score $\hat{\boldsymbol{y}}$ are automatically fused to regulate the impact of backdoor adjustment. 
We define a indicator function $\mathbf{I}_{a}$ that determines whether to include $\boldsymbol{c}_{\boldsymbol{a}}$ into the user intent $\boldsymbol{u}^{u}$ or not.
\begin{equation}
\mathbf{I}_{a}:=\left\{\begin{array}{ll}
\boldsymbol{c}_{\boldsymbol{a}} & \text { if } \tanh \left(\hat{\boldsymbol{y}}_{C}-\hat{\boldsymbol{y}}\right)>0 \\
\mathbf{1} & \text { if } \tanh \left(\hat{\boldsymbol{y}}_{C}-\hat{\boldsymbol{y}}\right)<0
\end{array}\right.
    \label{eq:ind}
\end{equation}
where $\tanh(\hat{\boldsymbol{y}}_{C} - \hat{\boldsymbol{y}})>0$ indicates that the backdoor adjustment leads a positive impact on recommendation result by considering the aspect $\boldsymbol{c}_{\boldsymbol{a}}$. Otherwise, $\boldsymbol{c}_{\boldsymbol{a}}$ leads a negative impact, which should be removed from user intent representation. 
Based on Eq.~\eqref{eq:ind}, the semantic-aware user intent representation can be refined as follows:
\begin{equation}
    \boldsymbol{e}_{\boldsymbol{a}}=\boldsymbol{u}^{u} \odot \mathbf{I}_{a}
\end{equation}
To define the unbiased loss function for observation $\hat{\boldsymbol{y}}_{ui}$, 
we aim to maximize the discrepancy between the final adjusted and the unadjusted representation $\boldsymbol{u}^{u}$ under the guidance of $\boldsymbol{e}_{\boldsymbol{a}}$, that is,
\begin{equation}\label{final_dis}
\small
\mathcal{L}_{d} =\underset{\bar{\theta}}{\arg \min }\sum_{(u, i, \boldsymbol{y}_{ui}) \in O} (\boldsymbol{y}_{ui}, f(\boldsymbol{u}, \boldsymbol{i}, \prod_a^{N} \boldsymbol{e}_{\boldsymbol{a}} ))
\end{equation}
where $\boldsymbol{u}$, $\boldsymbol{i}$ are the ID embedding for user $u$ and item $i$, $f(\cdot)$ is defined in Eq.~\eqref{calcu_y}, and $\boldsymbol{y}_{ui}$ is the ground-truth for user $u$ and item $i$.

\subsection{Optimization}
Our model ultimately has three loss functions, i.e., $\mathcal{L}_{d}$ of unbiased loss function for preference score estimation given in Eq.~\eqref{final_dis}, $\mathcal{L}_\theta$ of \emph{Heterogeneous Skip-Gram} model given in Eq.~\eqref{skip-gram}, and the BPR loss for preference score prediction given in Eq.~\eqref{BPR}.
To this end, the objective function of our CaDSI method could be derived as:
\begin{equation}\label{obejctive}
\begin{aligned}
\mathcal{L}= \lambda_{d}\mathcal{L}_{d}+\lambda_{\theta}\mathcal{L}_{\theta}+\lambda_z\mathcal{L}_{BPR}+\mathcal{R}(\Omega)
\end{aligned} 
\end{equation}
where $\Omega$ represents the trainable parameters and $\mathcal{R}(\cdot)$ is a squared $L_2$ norm regularization term on $\Omega$ to alleviate the overfitting problem.
$\lambda_{d}$, $\lambda_{\theta}$, $\lambda_z$ are trade-off hyper-parameters of the three separate loss functions respectively.
During the training, we optimize the objective function in Eq.~\eqref{obejctive} using gradient descent algorithm such as Adam~\cite{bushaev2018adam}.

\section{EXPERIMENTS}
To more thoroughly evaluate the proposed methd, 
experiments are conducted to answer the following research questions:
\begin{itemize}
    \item (\textbf{RQ1}) How confoundeing bias caused by the context information is manifested in real-world recommendation datasets?
    \item (\textbf{RQ2}) How does our model perform compared with state-of-the-art models for top-$N$ recommendation?
    \item (\textbf{RQ3}) How does key components in our model impact the recommendation performance? (i.e., disentanglement learning task, causal intervention)? How do hyper-parameters in our model impact recommendation performance?
    \item (\textbf{RQ4}) How does our model interprets  user intents for recommendation?
\end{itemize}
We first present the experimental settings for good reproducibility, followed by answering the above four research questions.

\subsection{Experimental Settings}

\subsubsection{Datasets}
We evaluate our model on three public accessible datasets for top-$N$ recommendation.
The statistics of the datasets are summarized in Table~\ref{tab_Data} and the selected meta paths for all data sets are shown in Table~\ref{mp_scheme}.
To ensure the quality of all the datasets, we use the core settings, i.e., we transform explicit ratings into implicit data, where each interaction between the user and item is marked as 0 or 1 indicating whether the user has rated the item or not; we retaining users and items with at least five interactions and each user has at least five friends for both of the datasets. 
In the training phase, each observed user-item interaction is treated as a positive instance, while we use negative sampling to randomly sample an unobserved item and pair it with the user as a negative instance.

\begin{table}[t]
\centering
\caption{\label{tab_Data} 
Statistics of three Datasets. Density of dataset is $\#Interactions/(\#Users \cdot\#Items)$, 
Avg.Degree of A is $\#Relation/\#A$, Avg.Degree of B is $\#Relation/\#B$.
}
\scriptsize
\begin{tabular}{|c||c|c|c|}
\hline
{\textbf{Dataset}} & {Node} & {Relation} & {Avg.Degree }\\
{(Density)} & {} & {A-B} & {of A/B}\\
\hline
\hline
\centering
& {\#User(U): 2,113} &{\#U-M: 855,598} & {\#U/M: 405.0/84.6}\\
&  {\#Movie(M): 10,109} & {\#U-U: 0} & {\#U/U: 0/0} \\
\texttt{MovieLens-HetRec} & {\#Actor(A): 38,044} & {\#M-A: 95,777} & {\#M/A: 9.5/2.5}\\
(4.0\%) & {\#Director(D): 4,031} & {\#M-D: 10,068} & {\#M/D: 1.0/2.5}\\
&{\#Country(C): 72} & {\#M-C: 10,109} & {\#M/C: 1.0/140.0}\\
&{\#Genre(G): 20} & {\#M-G: 20,670} & {\#M/G: 2.0/1033.5}\\
\hline

& {\#User(U): 13,024} & {\#U-Bo: 792,062} & {\#U/Bo: 60.8/35.4}\\
\texttt{Douban Book} &  {\#Book(Bo): 22,347} & {\#U-U: 169,150} & {\#U/U: 13.0/13.0}\\
& {\#Group(Gr): 2,936} & {\#U-Gr: 1,189,271} & {\#U/Gr: 91.3/405.1}\\
(0.27\%)&{\#Author(Au): 10,805} & {\#Bo-Au: 21,907} & {\#Bo/Au: 1.0/2.0}\\
&{\#Publisher(P): 1,815} & {\#Bo-P: 21,773} & {\#Bo/P: 1.0/12.0}\\
  &{\#Year(Y): 64} & {\#Bo-Y: 21,192} & {\#Bo/Y: 1.0/331.1}\\
\hline

  & {\#User(U): 13,367} & {\#U-M: 1,068,278} & {\#U-M: 79.9/84.3}\\
  &  {\#Movie(M): 12,677} & {\#U-U: 4,085} & {\#U/U: 1.7/1.8}\\
\texttt{Douban Movie} & {\#Group(Gr): 2,753} & {\#U-Gr: 570,047} & {\#U/Gr: 42.7/207.1}\\
(0.63\%) &{\#Actor(A): 6,311} & {\#M-A: 33,587} & {\#M-A: 2.9/5.3}\\
  &{\#Director(D): 2,449} & {\#M-D: 11,276} & {\#M/D: 1.1/4.6}\\
  &{\#Type(T): 38} & {\#M-T: 27,668} & {\#M/T: 2.2/728.1}\\
\end{tabular}
\end{table}

\begin{table}[t]
\centering
\caption{\label{mp_scheme} 
The selected meta paths for three datasets in our work.
}
\scriptsize
\begin{tabular}{|c||c|}
\hline
{Dataset} & {Meta path Schemes}\\
\hline
\hline
\centering
\texttt{MovieLens-HetRec} & {UMU, UMAMU, UMDMU, UMCMU, UMGMU}\\
  &{MUM, MAM, MDM, MCM, MGM}\\
\hline
\texttt{Douban Book} & {UBoU, UBoAuBoU, UBoPBoU, UBoYBoU, UBoAuBoU}\\
  &{BoUBo, BoPBo, BoYBo, BoAuBo}\\
\hline
\texttt{Douban Movie}& {UMU, UMDMU, UMAMU, UMTMU}\\
  &{MUM, MAM, MDM, MTM}\\
\hline
\end{tabular}
\end{table}

\subsubsection{Baselines}

To demonstrate the effectiveness, we compare our model with four classes of methods: (\Rmnum{1}) conventional entangled CF methods; (\Rmnum{2}) graph-based entangled recommendation methods; (\Rmnum{3}) HIN enhanced entangled recommendation methods, which model user-item interaction with rich context information as HIN; (\Rmnum{4}) disentangled recommendation methods, which disentangle user intents or item aspects with different mechanisms; (\Rmnum{5}) causal-based recommendation methods.
\begin{itemize}
    \item \textbf{NeuMF}~\cite{he2017neural} (\Rmnum{1}): This method combines deep neural networks with Matrix Factorization (MF) method for modeling the user-item interactions.
    \item \textbf{GC-MC}~\cite{berg2017graph} (\Rmnum{2}): The method organizes user behaviors as a graph, and employs one Graph Convolution Network (GCN) encoder to generate representations based on first-order connectivity.
    \item \textbf{NGCF}~\cite{wang2019neural} (\Rmnum{2}): This adopts three Graph Neural Network(GNN) layers to model at most third-order connectivity on the user-item interaction graph.
    \item \textbf{LightGCN}~\cite{he2020lightgcn} (\Rmnum{2}): This is a state-of-the-art graph-based recommendation method that learns user/item embeddings by linearly propagating them with neighborhood aggregation in the GCN component. 
    \item \textbf{IF-BPR}~\cite{yu2018adaptive} (\Rmnum{3}): This method leverages meta path based social relations derived from a HIN, and proposes a social recommendation method that can capture the similarity of users for top-$N$ recommendation. 
    \item \textbf{MCRec}~\cite{hu2018leveraging} (\Rmnum{3}): This method leverages meta path based context with co-attention mechanism for top-$N$ recommendation in HIN.
    \item \textbf{NeuACF}~\cite{han2018aspect} (\Rmnum{4}): This method disentangles multiple aspects of users and items with a deep neural network for recommendation in HIN.
    \item \textbf{MacridVAE}~\cite{ma2019learning} (\Rmnum{4}): This method disentangle user intents behind user behaviors, assuming that the co-existence of macro and micro latent factors affects user behaviors.
    \item \textbf{DGCF}~\cite{wang2020disentangled} (\Rmnum{4}): This is a state-of-the-art CF-based disentangled recommendation method, which disentangles latent factors of user intents by the neighbor routing and embedding propagation.
    \item \textbf{DICE}~\cite{10.1145/3442381.3449788} (\Rmnum{5}): This is a state-of-the-art causal-based recommendation method, which aims at disentangling users' interest by controlling the conformity bias using causal embedding.
\end{itemize}

\subsubsection{Evaluation Metrics}\label{metrics}
We adopt two popular metrics: Recall@$K$ and Normalized Discounted Cumulative Gain(NDCG)@$K$ to evaluate the top-$K$ recommendation performance of our model. $K$ is set as 20 by default.
In the inference phase, we view the historical items of a user in the test set as the positive, and evaluate how well these items are ranked higher than all unobserved ones. The average results w.r.t. the metrics over all users are reported.

\subsubsection{Parameter Settings}\label{para}

We implement all baseline models and our proposed CaDSI model on a Linux server with Tesla P100 PCI-E 16GB GPU. 
For a fair comparison, datasets for implementing all models are split as train/test/validate set with  a proportion of 80\%/10\%/10\% of the dataset, while we optimize all models with Adam~\cite{bushaev2018adam}.
For a fair comparison,  a grid search is conducted to choose the optimal parameter settings, e.g., dimension of user/item latent vector $k_{MF}$ for matrix factorization-based models 
and dimension of embedding vector $d$ for neural network-based models. 
The embedding size is initialized with the Xavier~\cite{glorot2010understanding} and searched in $\{8, 16, 32, 64, 128, 256\}$. The batch size and learning rate are searched in $\{32, 64, 128, 512, 1024\}$ and $\{0.0005,
0.001, 0.005, 0.01, 0.05, 0.1\}$, respectively.
The maximum epoch $N_{epoch}$ is set as 2000, an early stopping strategy is performed.
Moreover, we employ three hidden layers for the neural components of GC-MC
NGCF, LightGCN, MCRec, NeuACF, MacridVAE and DGCF.
The hyperparameter specifications of CaDSI are
set as: latent intents number $k$ as 4, iteration number of disentangling module $\boldsymbol{l}$ as 2, model depth of disentangling module $L$ as 2, iteration number of causal intervention $n$ as 140, and their influences are reported in Section~\ref{ablation}.

\subsection{Understanding Confounders (RQ1)}
\begin{figure}[htbp]
\centering
\begin{minipage}[t]{0.45\textwidth}
\centering
\includegraphics[width=0.48\textwidth]{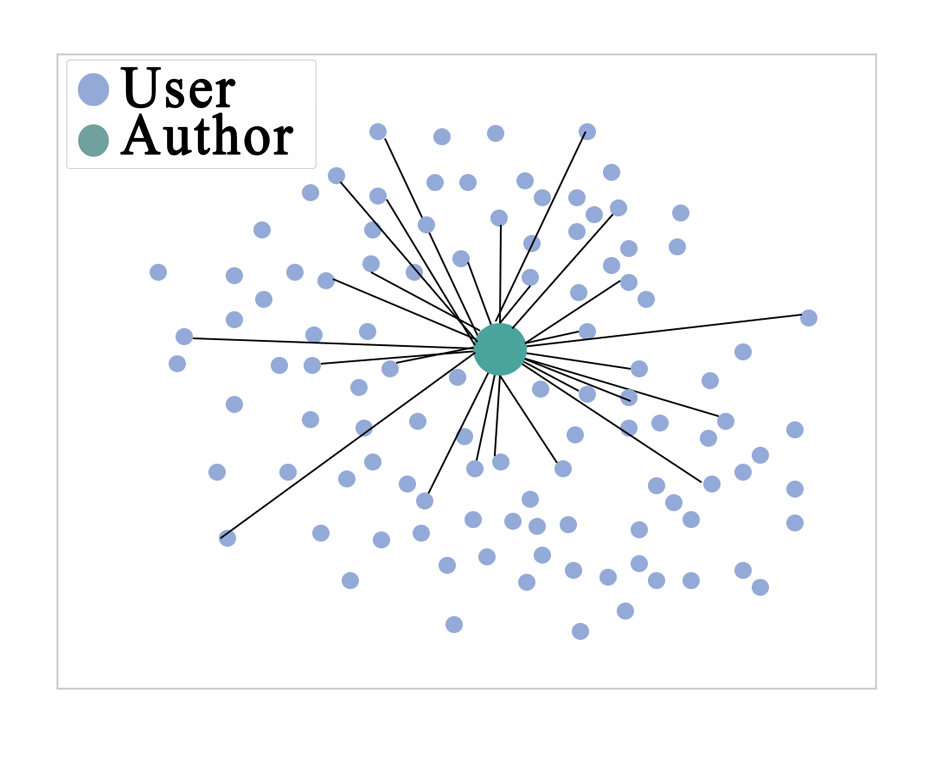}
\includegraphics[width=0.48\textwidth]{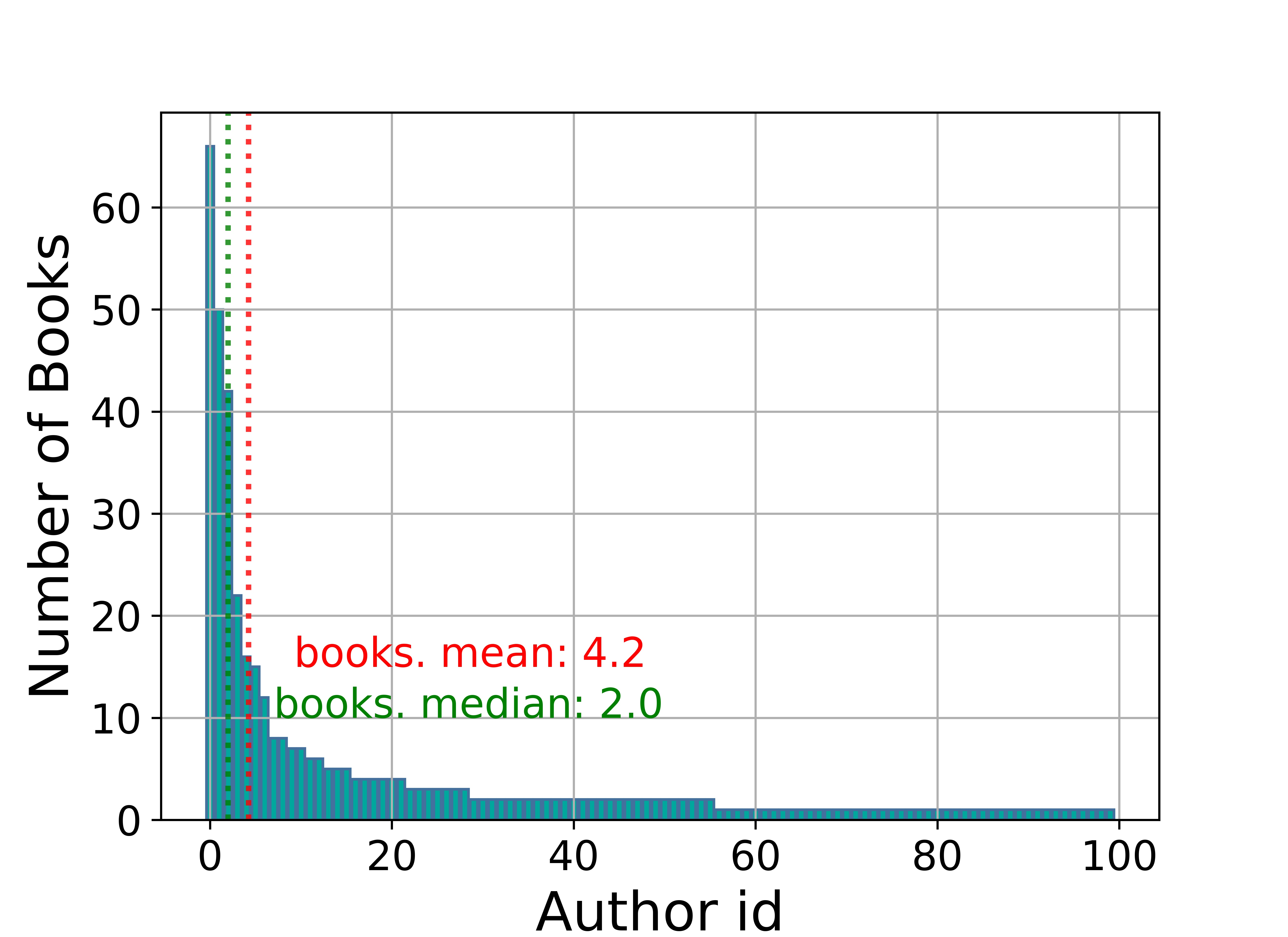}
\subcaption{(a)Distribution of $Book-Author$.}
\end{minipage}
\begin{minipage}[t]{0.45\textwidth}
\centering
\includegraphics[width=0.48\textwidth]{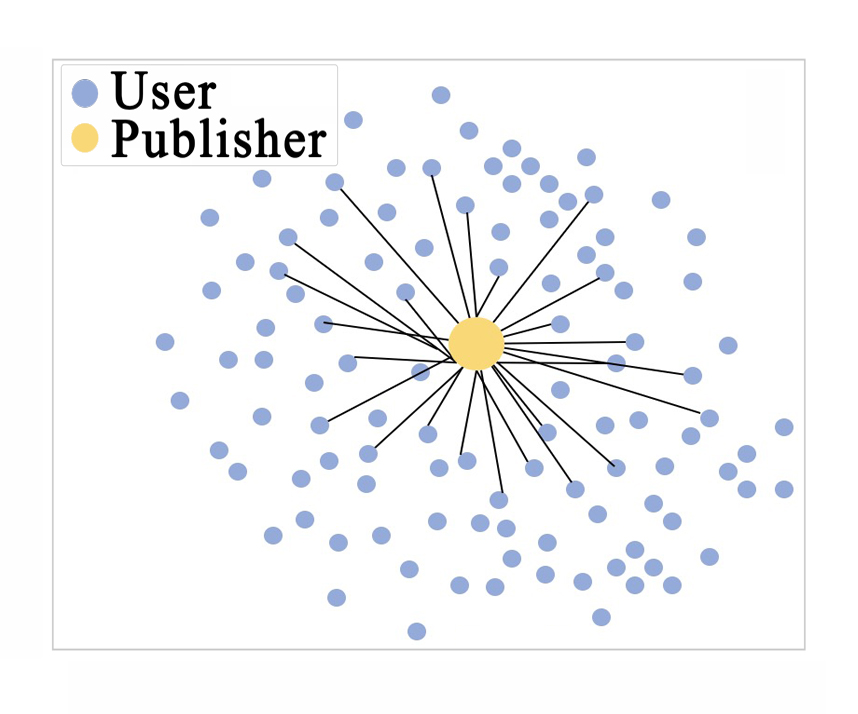}
\includegraphics[width=0.48\textwidth]{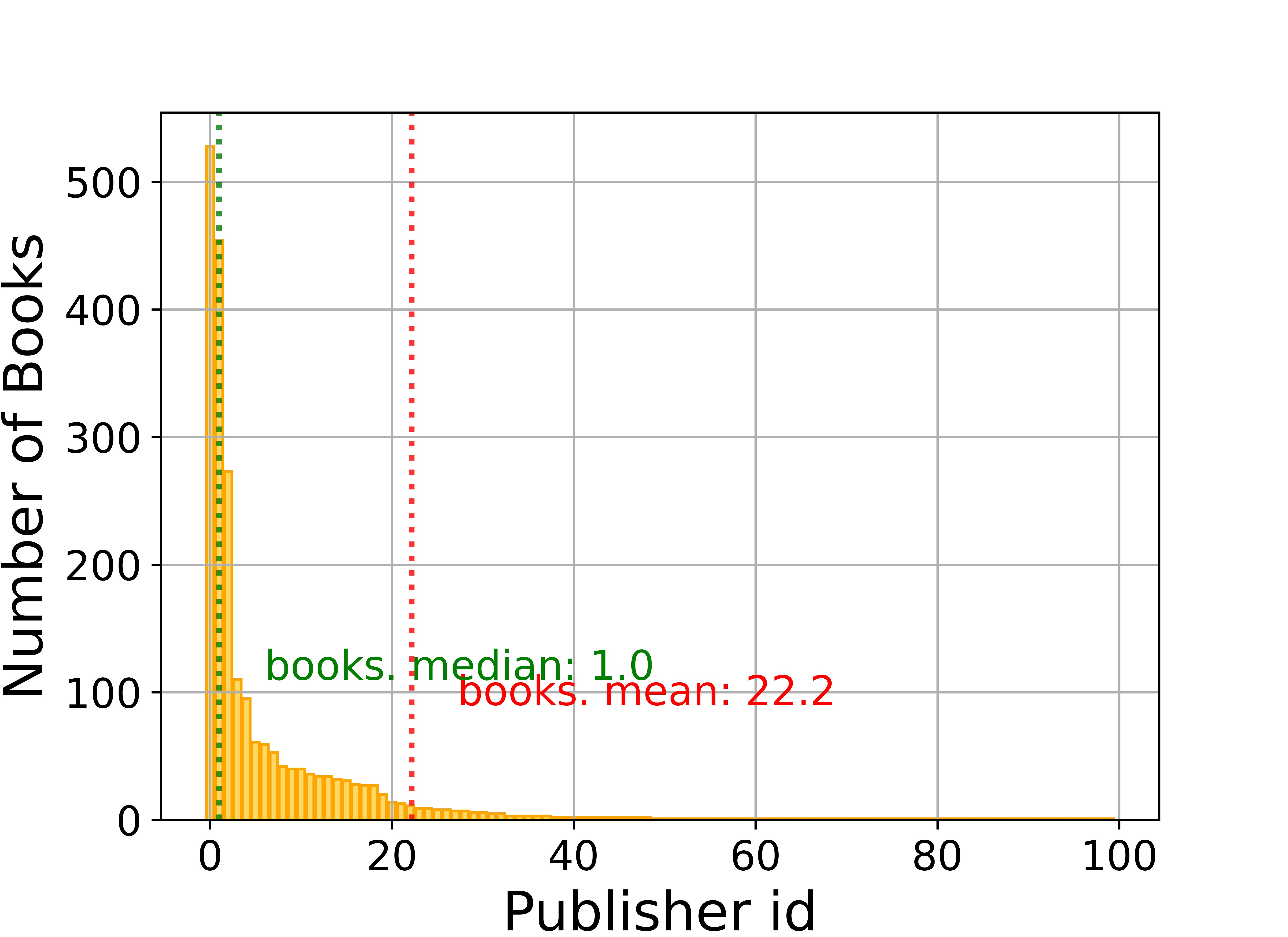}
\subcaption{(b) Distribution of $Book-Publisher$.}
\end{minipage}
\begin{minipage}[t]{0.45\textwidth}
\centering
\includegraphics[width=0.48\textwidth]{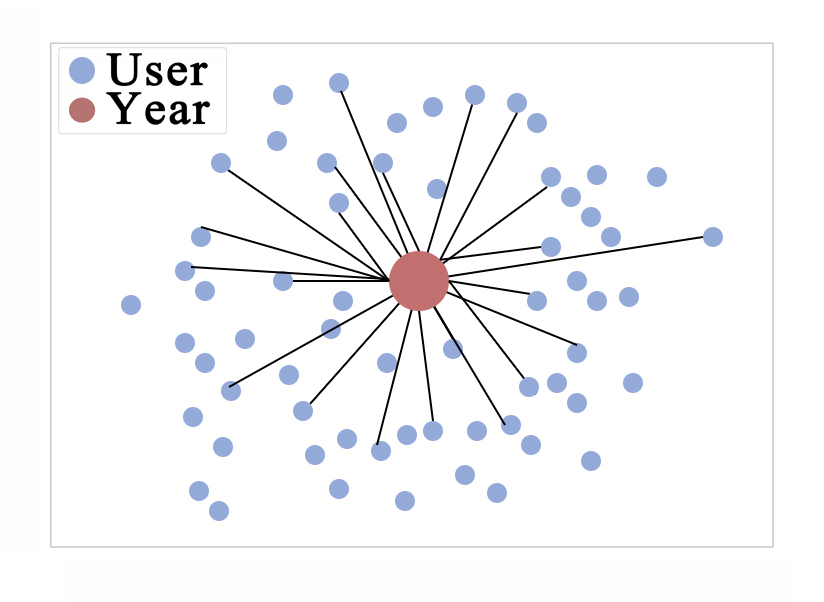}
\includegraphics[width=0.48\textwidth]{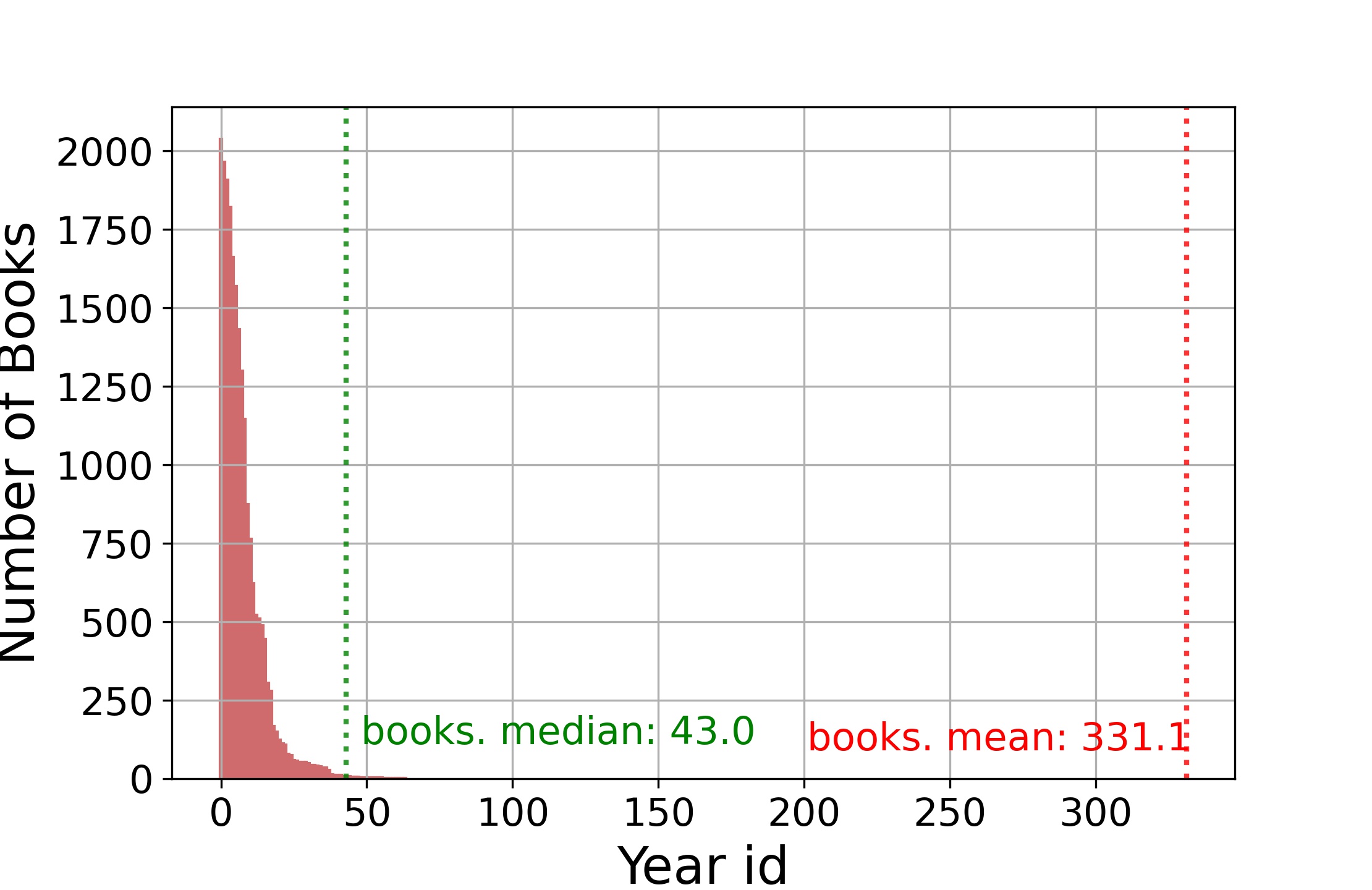}
\subcaption{(c) Distribution of $Book-Year$.}
\end{minipage}
\caption{The distributions of $Book-Author$, $Book-Publisher$ and $Book-Year$ of \texttt{Douban Book} dataset.}
\label{fig:rq1}
\end{figure}

We initially conduct an experiment to understand to what extent the confounding bias exists in meta paths of real-world recommendation datasets.
To this end, we aim to investigate the distribution of nodes among the same meta path.
Intuitively, an unbiased HIN-based recommendation method should expect that, for a specific attribute, each user/item should hold an equal number of this attribute (i.e, interactions between nodes and attributes are likely to be evenly distributed).
Thus, we investigate the confounding bias by analyzing the statistics of node-attribute interactions of meta paths in \texttt{Douban Book}. 
We randomly sample $n=100$ books from \texttt{Douban Book} and extract their \emph{Author}, \emph{Publisher}, and \emph{Year} attributes. By counting the connections between books and their attributes whose type belongs to \emph{Author}, \emph{Publisher}, and \emph{Year}, respectively, we have the statistical results shown in Figure~\ref{fig:rq1}. The connected graphs in the left part of Figure~\ref{fig:rq1} depict whether the book $i$ connected with the selected attribute.
The figures in the right part of Figure~\ref{fig:rq1} shows the distributions of connected book numbers by a certain attribute, where the $y$-axis denotes the total amount of the connected books.

Apparently, attributes and books exhibit an unevenly distribution regarding their interactions:
the attributes in dataset are partially observed, leaving a larger number of attributes to be unobserved. 
For example, for \emph{Book-Author} meta path, there are a lot of books that do not connect with any node whose type is \emph{Author}, which means lots of author attributes of books are missing. In conventional recommendation methods, the missing pattern of such attributes is ignored by either regarding them as outliers and padding them with random values, or treating the missing attributes as negative feedback. 
Such measure would preserve the confoundings bought by node attributes, degrading the recommendation ultimately.

Another finding is that, the distribution for book-attribute connection numbers is significantly skewed, it displays a long-tail phenomenon: the green vertical line separates the top 50\% of connection numbers by popularity - these connections outweigh another 50\% long tail connections to the right.
For instance, in Figure~\ref{fig:rq1} (a), authors in the \emph{Book-Author} relation cumulatively connect with 90\% more books than the long tail authors to the right.
For \emph{Book-Publisher} meta path, ideally, \emph{Book-Publisher} has the one-to-one relation from book to the publisher, while every publisher has published an equal number of books. 
However, some publishers have published at most $510$ books, while more than 90\% of publishers only published fewer than $10$ books.
Such long-tail distribution can bias the users' interest on item aspects.
i.e., recommendation methods tend to recommend those items that have the most frequent attribute, while users can only be exposed to those that are recommended. 

\subsection{Performance Comparison (RQ2)}
We compare the top-$K$ recommendation performance of CaDSI with ten recommendation baselines on three datasets: \texttt{MovieLens-HetRec},
\texttt{Douban Book} and 
\texttt{Douban Movie}. 
Table~\ref{tab:overall-performance} demonstrates the performance comparison and we have the following observations:

\begin{table*}[t]
    \caption{Overall Performance Comparison: bold numbers are the improvement percentages; the best results are marked with $^{*}$, strongest baselines are marked  with underline.
    }
    \centering
    \label{tab:overall-performance}
    \begin{tabular}{l|cccc|cccc|cccc}
    \hline
     & \multicolumn{4}{c|}{ \texttt{MovieLens-HetRec}} & \multicolumn{4}{c|}{ \texttt{Douban Book} } & \multicolumn{4}{c}{\texttt{Douban Movie} } \\
     & Recall@20 & NDCG@20 & Recall@40 & NDCG@40 & Recall@20 & NDCG@20 & Recall@40 & NDCG@40 & Recall@20 & NDCG@20 & Recall@40 & NDCG@40\\ \hline\hline
    NeuMF & 0.0434 & 0.0557 & 0.0665 &\underline{0.0709} & 0.0339 &  0.0391 & 0.0641 & 0.0682 &0.0460  &0.0417 & 0.0708 & 0.0611\\
    \hline
    GC-MC & 0.0336 & 0.0404 & 0.0653 & 0.0584 & 0.0458 & 0.0402 & 0.0675 & 0.0643 &0.0448 & 0.0461 & 0.0602 & 0.0622\\ 
    NGCF & 0.0365 & 0.0508 & 0.0699 & 0.0615 & 0.0252 & 0.0301 & 0.0707 & 0.0691 & 0.0475 & 0.0498 & 0.0689 & \underline{0.0642}\\ 
    LightGCN & 0.0466 & 0.0155 & 0.0615 & 0.0498 & 0.0201 & 0.0225 & 0.0531 & 0.0568 & 0.0294 & 0.0331 & 0.0499 & 0.0578\\
    \hline
    IF-BPR & 0.0546 & 0.0510 & 0.0727 & 0.0689 & 0.0396 & 0.0463 & 0.0628 & 0.0601 & 0.0483 & 0.0501 & 0.0652 & 0.0603\\
    MCRec & 0.0352 & 0.0195 & 0.0680 & 0.0677 & 0.0165 & 0.0294 & 0.0481 & 0.0507 & 0.0281 & 0.0336 & 0.0618 & 0.0629\\
    \hline
    NeuACF & 0.0236 & 0.0308 & 0.0556 & 0.0684 & 0.0298 & 0.0201 & 0.0601 & 0.0579  &0.0351 & 0.0438 & 0.0571 & 0.0623\\
    MacridVAE & 0.0454 & 0.0290 & 0.0661 & 0.0592 & 0.0309 & 0.0425 & 0.0691 & 0.0645 & 0.0489 & 0.0441 & 0.0729 & 0.0616 \\
    DGCF & 0.0229 & \underline{0.0589} & 0.0532 & 0.0708 & 0.0431 & 0.0502 & 0.0649 & 0.0663 & 0.0416 & \underline{0.0527} & 0.0702 & 0.0628 \\
    \hline
    DICE  & \underline{0.0549}  & 0.0499  & \underline{0.0740} & 0.0703 & \underline{0.0577} & \underline{0.0608} & \underline{0.0820} & \underline{0.0799} &  \underline{0.0513} & 0.0389 & \underline{0.0811} & 0.0636\\
    \hline
    Our model & $0.0678^{*}$ & $0.0659^{*}$ & $0.0765^{*}$ & $0.0736^{*}$ & $0.0708^{*}$ & $0.0722^{*}$ & $0.1466^{*}$ & $0.1194^{*}$ & $0.0583^{*}$ & $0.0547^{*}$ & $0.0918^{*}$ & $0.0647^{*}$\\ 
    \hline\hline
    \%improv. &\textbf{23.5\%} & \textbf{11.9\%} & \textbf{3.4\%} & \textbf{3.8\%} & \textbf{22.7\%} & \textbf{18.8\%} & \textbf{78.8\%} & \textbf{49.4\%} & \textbf{13.6\%}  & \textbf{3.8\%} & \textbf{13.2\%} & \textbf{0.8\%}\\
    \hline
    \end{tabular}
\end{table*}

\begin{itemize}
\item Our CaDSI consistently yields the best performance among all methods on three datasets.
In particular,
CaDSI improves over the strongest baselines w.r.t. Recall@20 by 23.5\%, 22.7\%, 13.6\% , NDCG@20 by 11.9\%, 18.8\%, 3.8\%, Recall@40 by 3.4\%, 78.8\%, 13.2\% and NDCG@40 by 3.8\%, 49.4\%, 0.8\%
on \texttt{MovieLens-HetRec}, \texttt{Douban Book} and \texttt{Douban Movie} respectively.  
CaDSI outperforms all baseline methods on top-$K$ recommendation task, which validates that the semantics-aware user intents representation can enhance the recommendation performance.

\item 
In virtue of user-item interaction graph and meta paths,  GNN-based (GC-MC, NGCF and LightGCN) and HIN-based (IF-BPR, MCRec) recommendation methods can achieve better performance than conventional MF methods (NeuMF) in most cases. 
However, they ignore controlling the bias existing in the context information.
On the contrary, our CaDSI adoptes a principled causal inference way to easing such confounding bias. So it outperforms GNN-based and HIN-based recommendation method on both of the datasets.
For instance, our CaDSI outperforms the most competitive HIN-based recommender IF-BPR w.r.t. Recall@20/Recall@40 by 24.2\%/5.2\% and NDCG@20/NDCG@40 by 29.2\%/6.8\% on \texttt{MovieLens-HetRec}. 

\item By performing unbiased disentanglement via semantics context, our CaDSI can infer user's potential interests of items. However, those user interests could not be well inferred from other disentangled recommendation methods (NeuACF, MacridVAE and DGCF).

\item 
Among the 
GNN-based (GC-MC, NGCF and LightGCN) and HIN-based recommenders (IF-BPR, MCRec ) recommenders and disentangled recommenders (NeuACF, MacridVAE and DGCF),
causal-based disentangled method (DICE) serves as the strongest baseline in most cases. 
This justifies the effectiveness of easing the counfounding bias in context information when estimating disentangled users' interests.
However, DICE performs worse than our CaDSI, as it ignores rich semantics information in HIN, and fails to ingest semantics aspects when disentangling user interests.

\item
From movie recommendation datasets, we can find that the improvements on \texttt{MovieLens-HetRec} is bigger than that on \texttt{Douban Movie}. This is reasonable since \texttt{Douban Movie} is much more sparser than \texttt{MovieLens-HetRec} 
with sparsity rate 0.63\% vs. 4.0\%, respectively. 
However, our CaDSI has better performance than all baselines on \texttt{Douban Movie}, because it achieves unbiased evaluation on high-order connectivity and rich semantics.
This indicates that CaDSI 
is robust to the very sparse dataset.
\end{itemize}

\subsection{Study of CaDSI\textbf{ (RQ3)}}\label{ablation}
Ablation studies on CaDSI are also conducted to investigate the
rationality and effectiveness. 
Specifically, we first attempt to exploit how the \emph{disentangled learning} and \emph{causal intervention} affect our performance.
Moreover, the stability of our approach's performance on top-$K$ recommendation is validated as well.

We have one fixed parameter $n=140$ (\textit{cf.} Eq.~\eqref{backdoor}) which denotes the total number of causal intervention times. Three important hyperparameters $k$ (\textit{cf.} Eq.~\eqref{u_initit}), $L$ (\textit{cf.} Eq.~\eqref{layer}) and $K$ (\textit{cf.} Section~\ref{metrics})
correspond to: the number of latent factors of user intents, the number of graph disentangling layers and the number of items in top-$K$ recommendation list, respectively. 
Based on the hyperparameter setup in Section~\ref{para}, for all questions listed above, we vary the value of one parameter while keeping the others unchanged. 

\subsubsection{Effect of Disentanglement Learning}
\begin{figure}[htbp]
\centering
\begin{minipage}[t]{0.5\textwidth}
\centering
\includegraphics[width=0.32\textwidth]{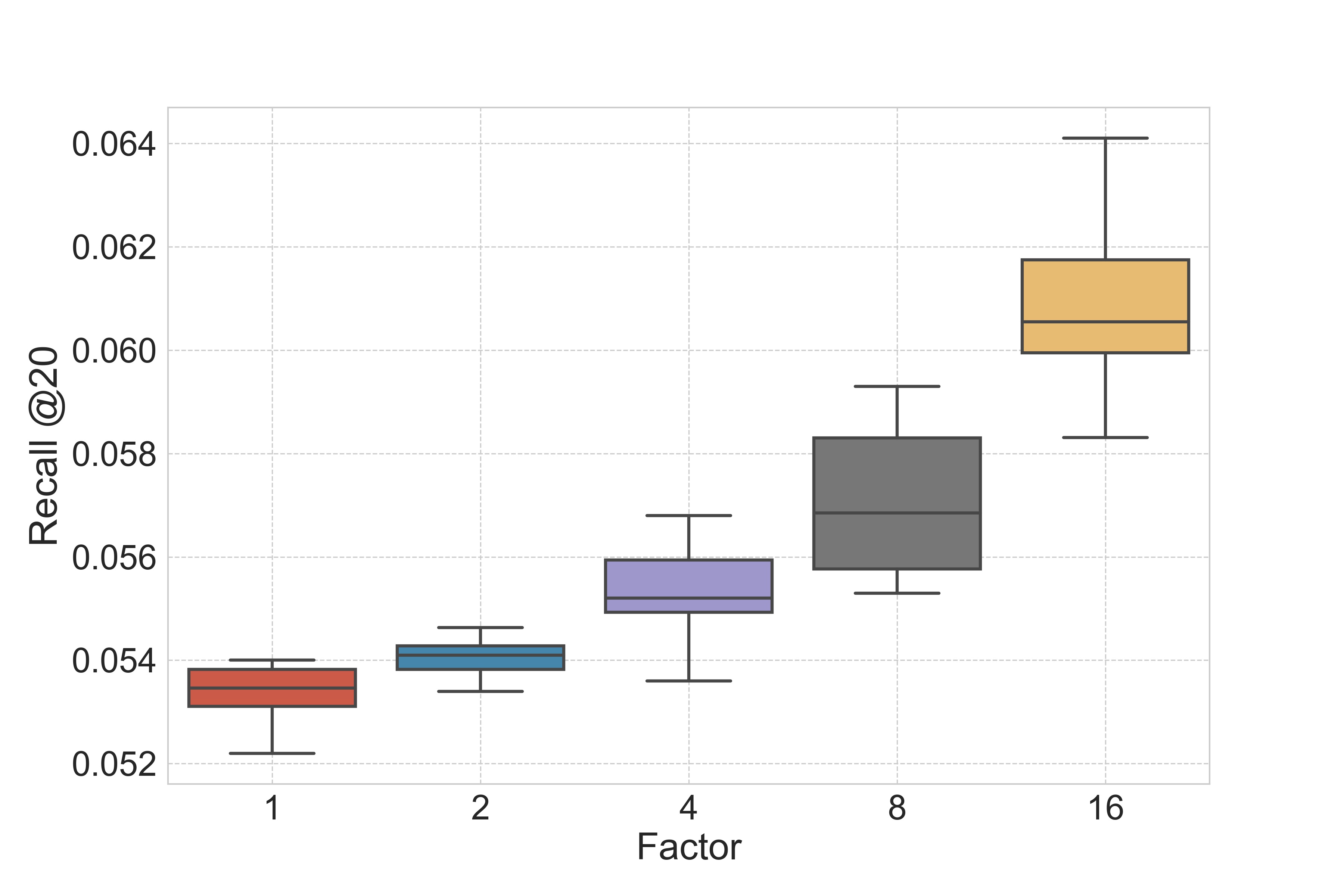}
\includegraphics[width=0.32\textwidth]{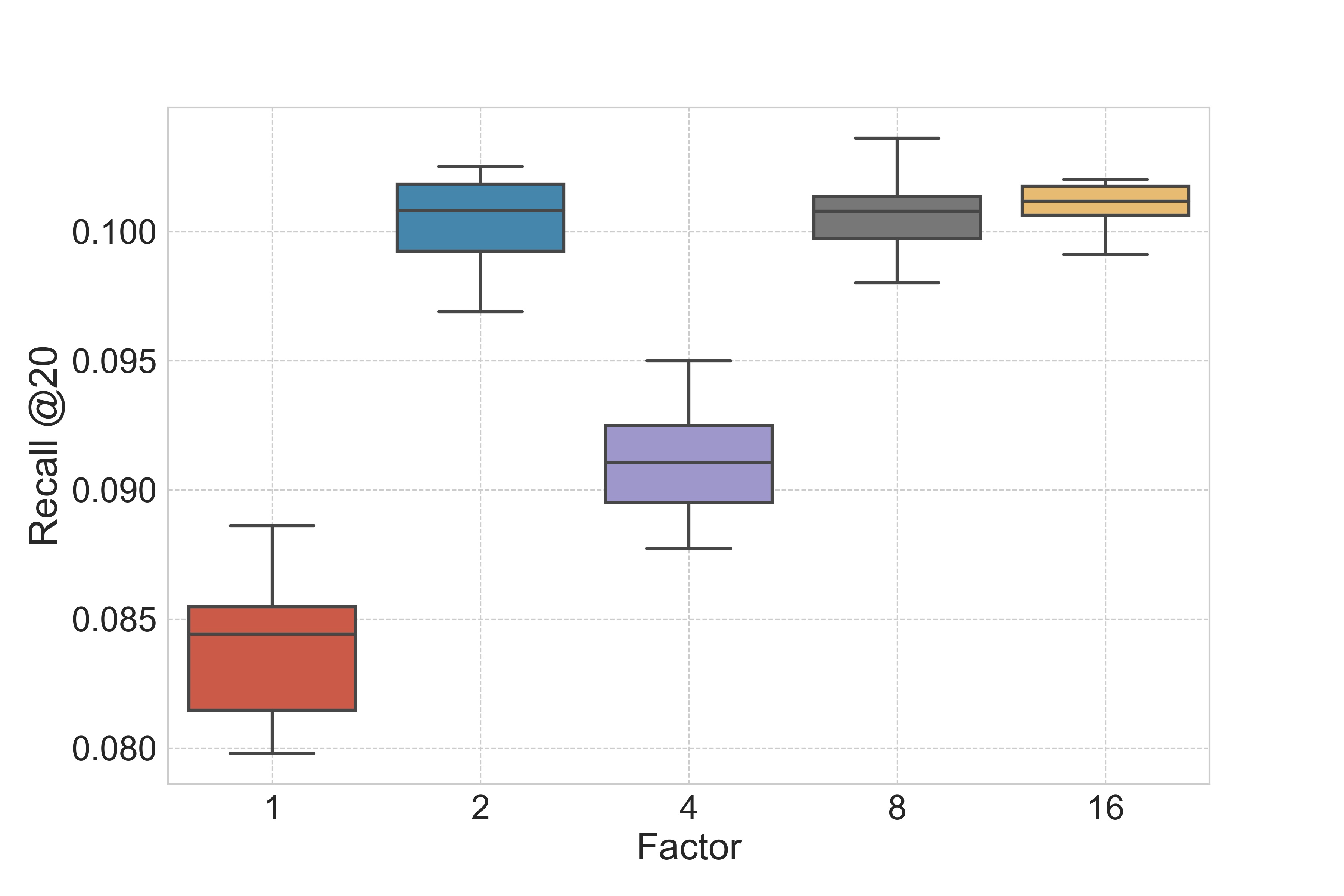}
\includegraphics[width=0.32\textwidth]{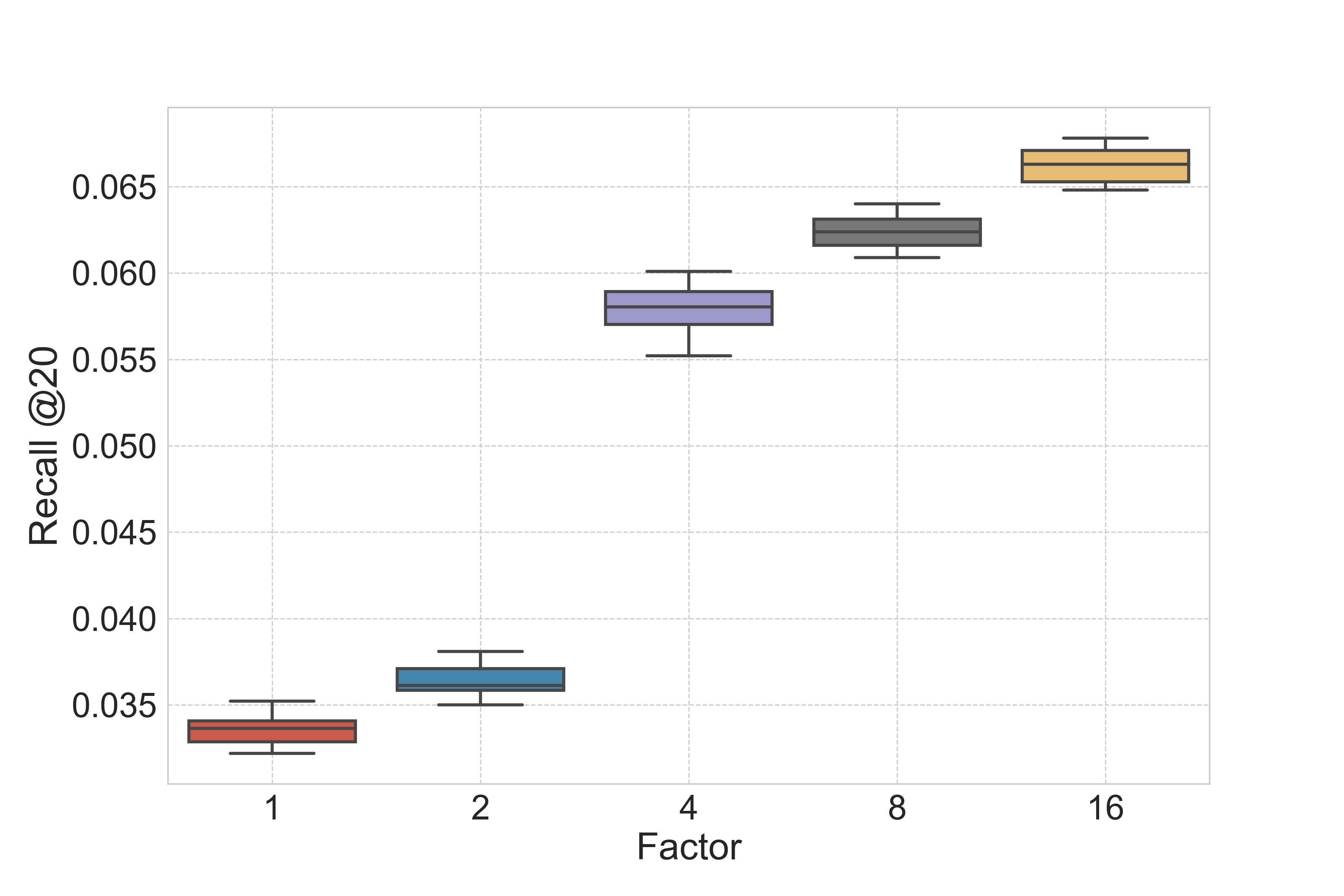}
\subcaption{(a) Recall@20 on \texttt{MovieLens-HetRec}, \texttt{Douban Book} and \texttt{Douban Movie}.}
\end{minipage}
\begin{minipage}[t]{0.5\textwidth}
\centering
\includegraphics[width=0.32\textwidth]{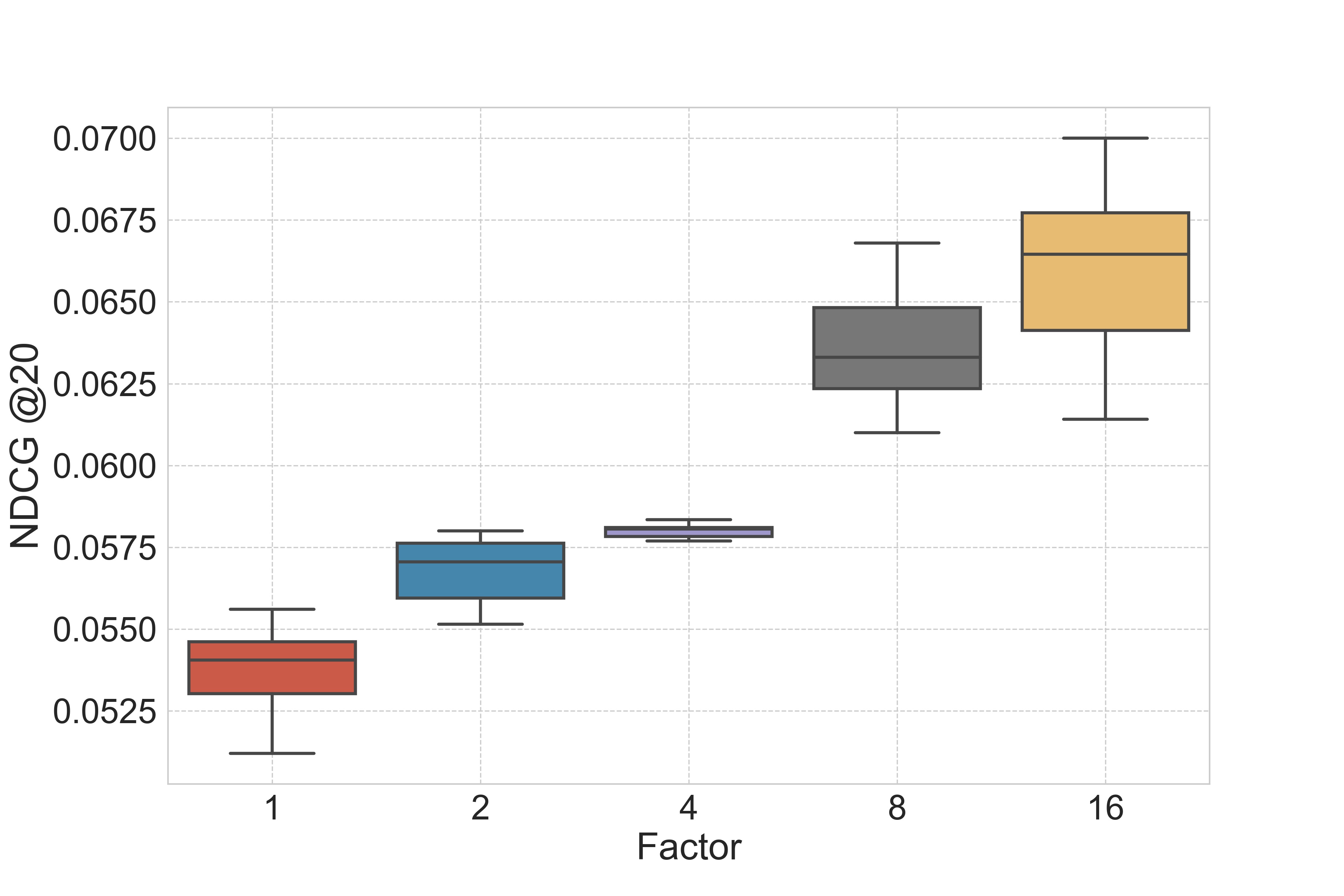}
\includegraphics[width=0.32\textwidth]{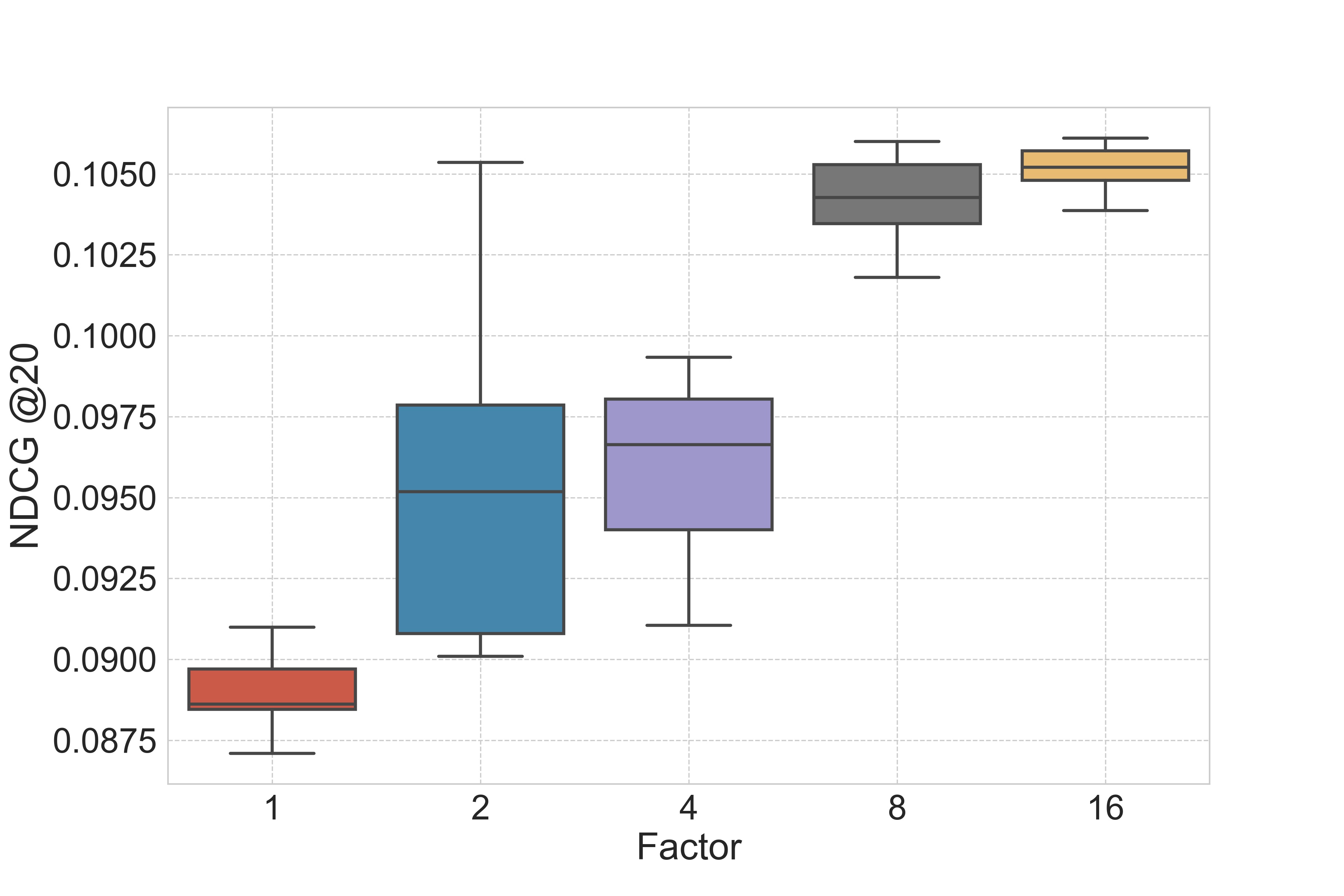}
\includegraphics[width=0.32\textwidth]{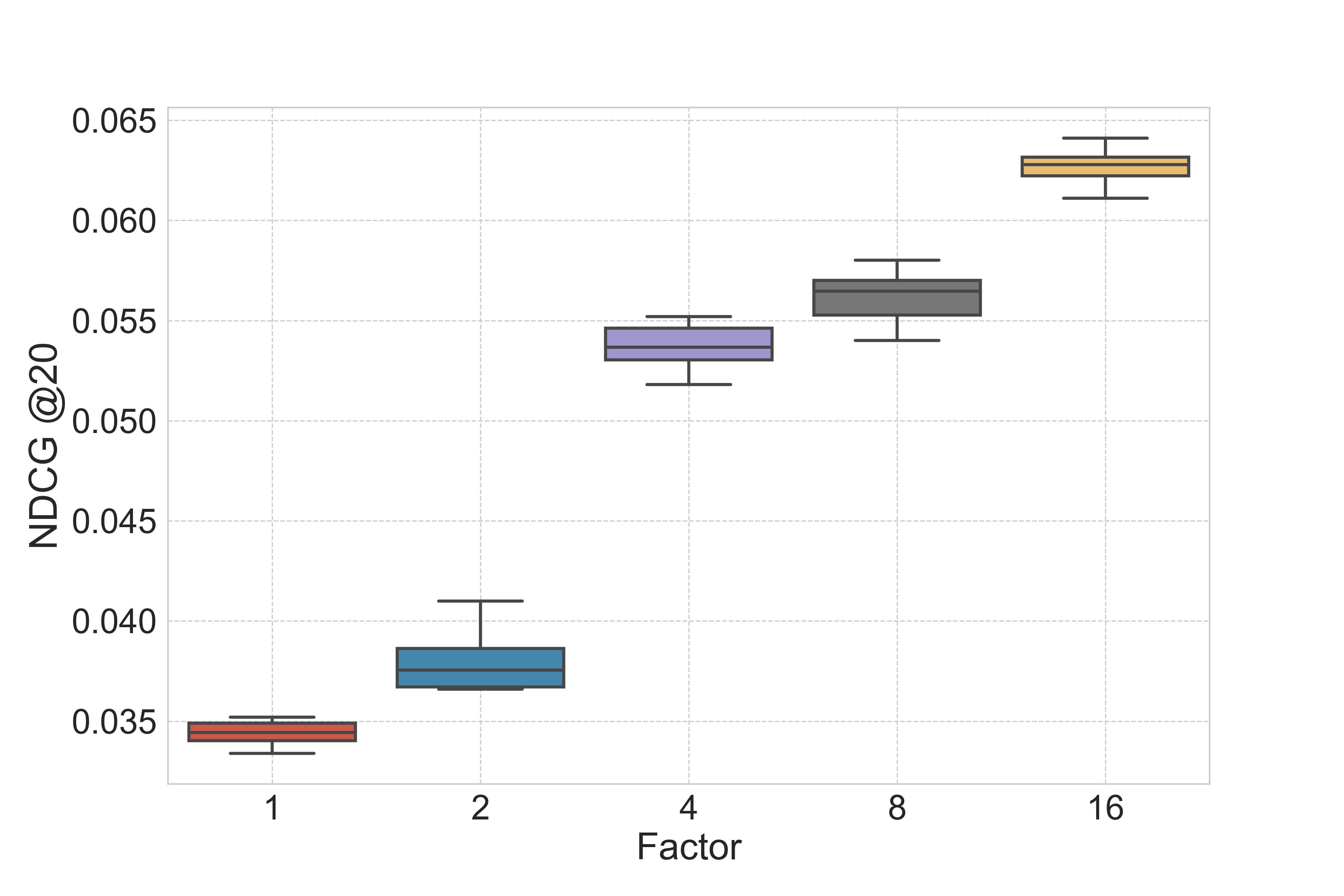}
\subcaption{(b) NDCG@20 on \texttt{MovieLens-HetRec}, \texttt{Douban Book} and \texttt{Douban Movie}.}
\end{minipage}
\caption{The recommendation performance comparison under different latent user intent factors.}
\label{fig:intent}
\end{figure}

The intent number $k$ controls the total amount of user intents considered in our model, larger $k$ stands for more fine-grained disentangled user intents. 
To study the influence, we vary $k$ in the range of $\{1,2,4,8,16\}$ and show the corresponding performance comparison on \texttt{MovieLens-HetRec}
\texttt{Douban Book}, 
\texttt{Douban Movie} in Figure~\ref{fig:intent}. We have several observations.

\begin{itemize}
    \item Increasing the intent number from 1 to 16 can significantly enhances the performance, while CaDSI performs the worst when $k=1$. 
    This indicates learning the disentanglement of user intents is effective to capture the real user preferences towards items instead of coupling all preference together.

    \item The variations diverse across different datasets. For \texttt{MovieLens-HetRec} and \texttt{Douban Movie}, the performance of CaDSI increase steadily as the $K$ value increases from 1 to 16, while the performance drops when $k$ is set from 2 to 4 on \texttt{Douban Book}. 
    One possible reason is that CaDSI should balance between too fine-grained disentangled intents and the adjustment from causal intervention, such balancing learning is more obvious when dataset size is lager. 
\end{itemize}

\subsubsection{Effect of Causal Intervention}
To investigate whether CaDSI can get benefit from causal intervention, we study the performance of CaDSI by varying the iterations of causal intervention.
Figure~\ref{fig:ablation_inter} summarizes
the experimental results w.r.t. \texttt{MovieLens-HetRec}
\texttt{Douban Book}, 
\texttt{Douban Movie}
and we have the following observations:

\begin{figure}[htbp]
\centering
\begin{minipage}[t]{0.15\textwidth}
\centering
\includegraphics[width=\textwidth]{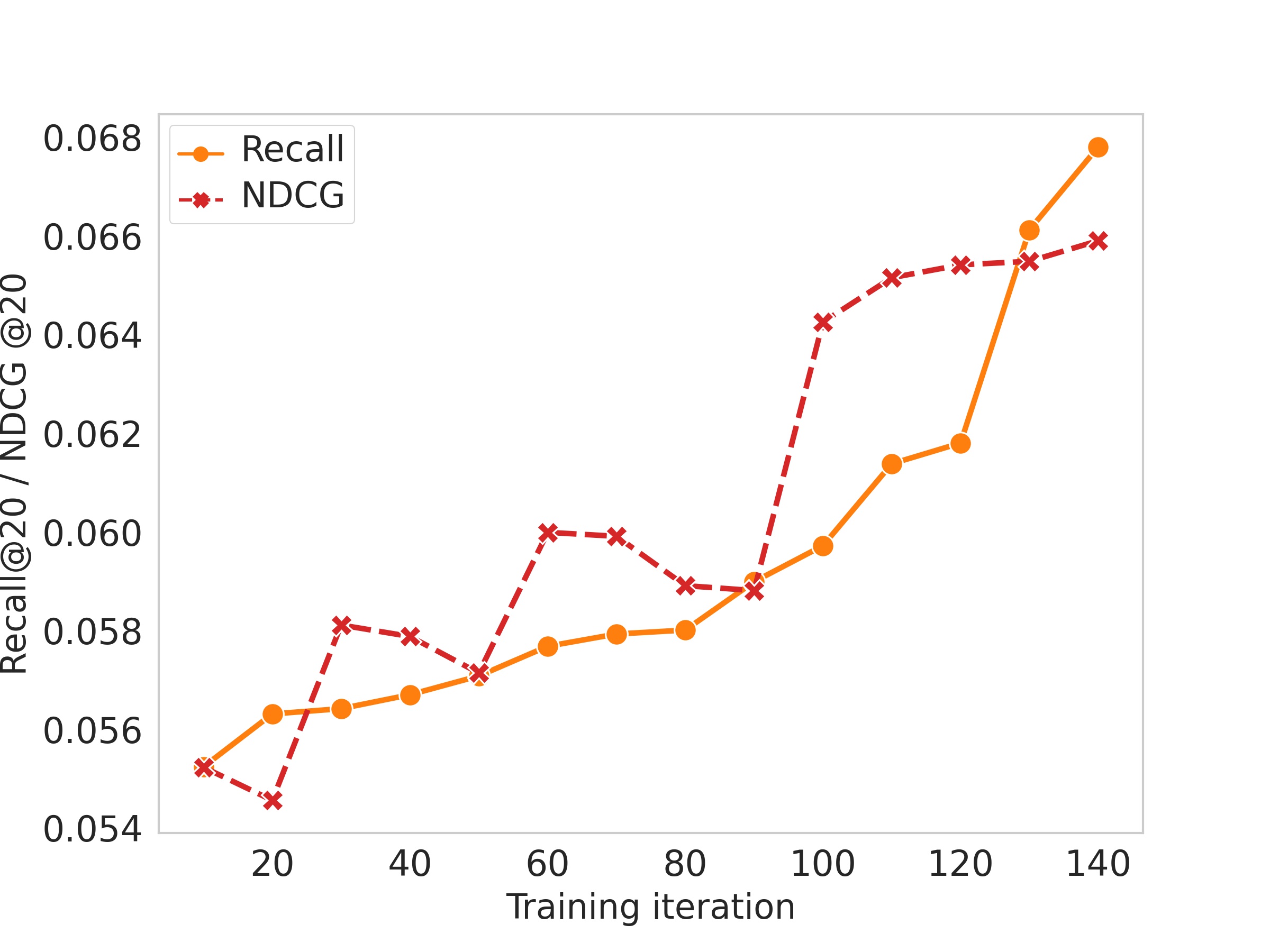}
\subcaption{(a) \texttt{MovieLens-HetRec}.}
\end{minipage}
\begin{minipage}[t]{0.15\textwidth}
\centering
\includegraphics[width=\textwidth]{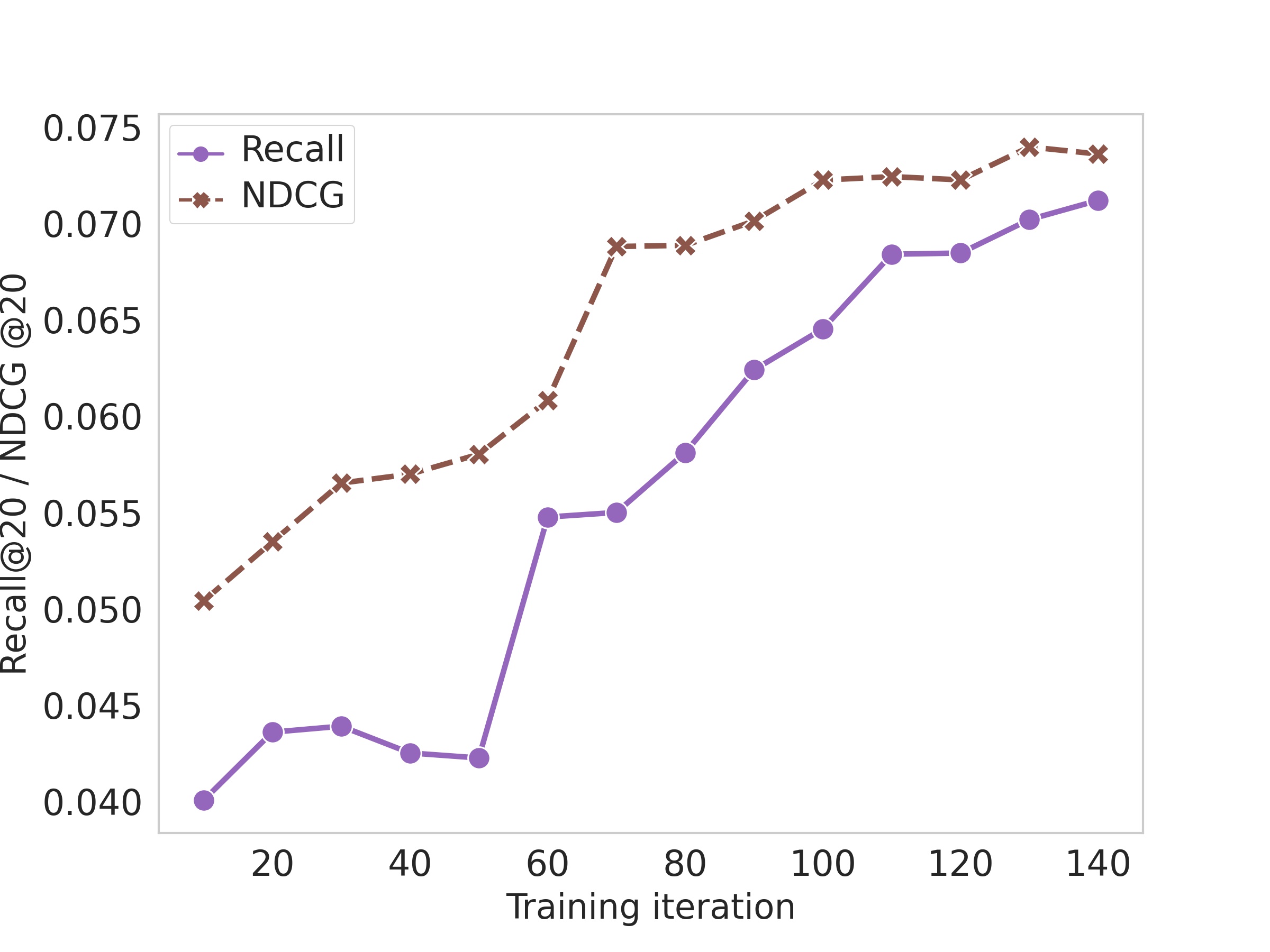}
\subcaption{(b) \texttt{Douban Book}.}
\end{minipage}
\begin{minipage}[t]{0.15\textwidth}
\centering
\includegraphics[width=\textwidth]{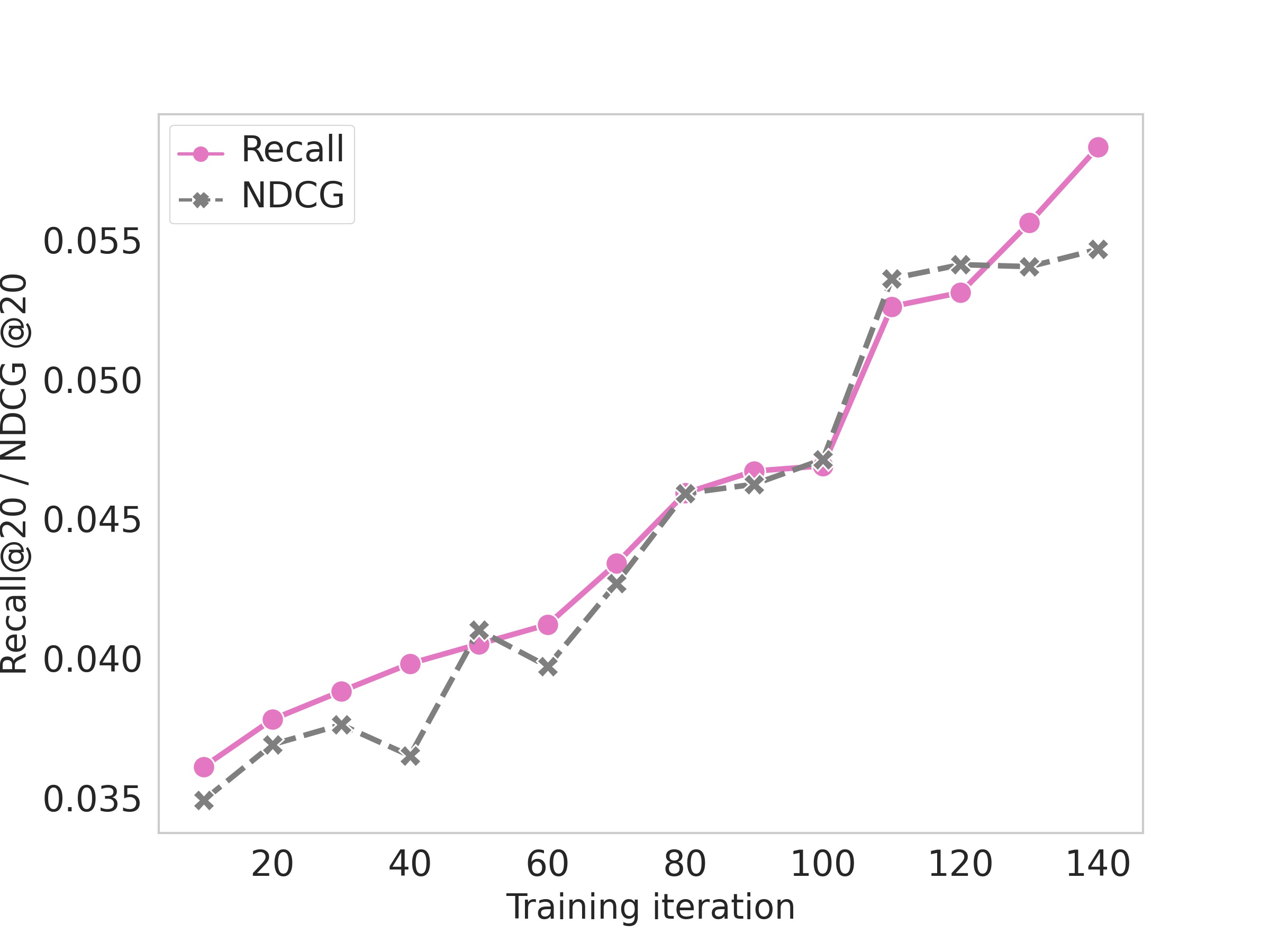}
\subcaption{(c) \texttt{Douban Movie}.}
\end{minipage}
\caption{Impact of causal intervention on the recommendation performance of our CaDSI along with iterations.}
\label{fig:ablation_inter}
\end{figure}

\begin{itemize}
    \item Clearly, the causal intervention mechanism renders our CaDSI a better recommendation performance: more iterations of causal intervention lead to the better recommendation performance before saturation on all datasets, e.g., the Recall@20 and NDCG@20 values generally increase along with training iterations in Figure~\ref{fig:ablation_inter}.
    \item When training iterations reach to 130, 70 and 110 for \texttt{MovieLens-HetRec}, \texttt{Douban Book} and \texttt{Douban Movie}, respectively, the performance becomes relatively stable. 
    Moreover, \texttt{Douban Book} requires less iteration times than the other two datasets.
    Intuitively, purchasing books is a much more simple behavior than choosing movies. Thus the user intents on book aspects are less diverse, leading to a quick convergence to the optimal interventional representations.
    \item  Some fluctuations appear in the iteration process, especially on \texttt{MovieLens-HetRec} dataset.
    The size of \texttt{MovieLens-HetRec} is much smaller than the other two datasets, 
    thus leading to an instability to intervention process due to the data sparsity. 
    However, when carrying more iterations, small-size datasets such as \texttt{MovieLens-HetRec} can also yield satisfying results.
\end{itemize}

\subsubsection{Effect of Multi-order Connectivity}
Since CaDSI is benefited from the higher-order connectivity between complex interactions and context information,
we investigate how connectivity degrees affect  CaDSI.
Specifically, we search the graph disentangling layer number $L$ in the range of $\{1,2,3\}$, which correspond to first-order connectivity, second-order connectivity and third-order connectivity, respectively. 
We show the performance comparison  in Table~\ref{tab:ablation_layer} and below are our observations. 

\begin{table}[]
\caption{Impact of multi-order connectivity (i.e., graph propagation layer number $L$) on \texttt{MovieLens-HetRec}, \texttt{Douban Book} and \texttt{Douban Movie}.}
\label{tab:ablation_layer}
\resizebox{0.5\textwidth}{!}{%
\begin{tabular}{|l|ll|ll|ll|}
\hline
 & \multicolumn{2}{l|}{\texttt{MovieLens-HetRec}} & \multicolumn{2}{l|}{\texttt{Douban Book}} & \multicolumn{2}{l|}{\texttt{Douban Movie}} \\ \hline
Layer number & Recall & NDCG   & Recall  & NDCG    & Recall & NDCG \\ \hline
1            & 0.0505 & 0.0521 & 0.0651 & 0.0684 & 0.0583 & 0.0546     \\ \hline
2            & 0.0672 & 0.0683 & 0.0712  & 0.0736  & 0.0596 & 0.0570     \\ \hline
3            & 0.0611 & 0.0624 & 0.0682  & 0.0701  &  0.0562 & 0.0573     \\ \hline
\end{tabular}%
}
\end{table}

\begin{itemize}
    \item  More graph disentangling layers will collect more information form multi-hop neighbors from a holistic user-item interaction graph. Clearly,  the performance of our CaDSI with layer number $L=2$ is better than that with $L=1$, since the second-order connectivity can capture significant collaborative signals with respect to users and items. 
    \item When stacking more than 2 layers, the influence of multi-hop neighbors is small and the recommendation performance is degraded. 
    This is reasonable since the informative signals of user-item interactions might introduce additional noises to the representation learning. 
    This again emphasizes the importance of controlling the bias bought by context information. 
\end{itemize}

\subsubsection{Top-$K$ Recommendation Performance}
\begin{figure}[htbp]
\centering
\begin{minipage}[t]{0.15\textwidth}
\centering
\includegraphics[width=\textwidth]{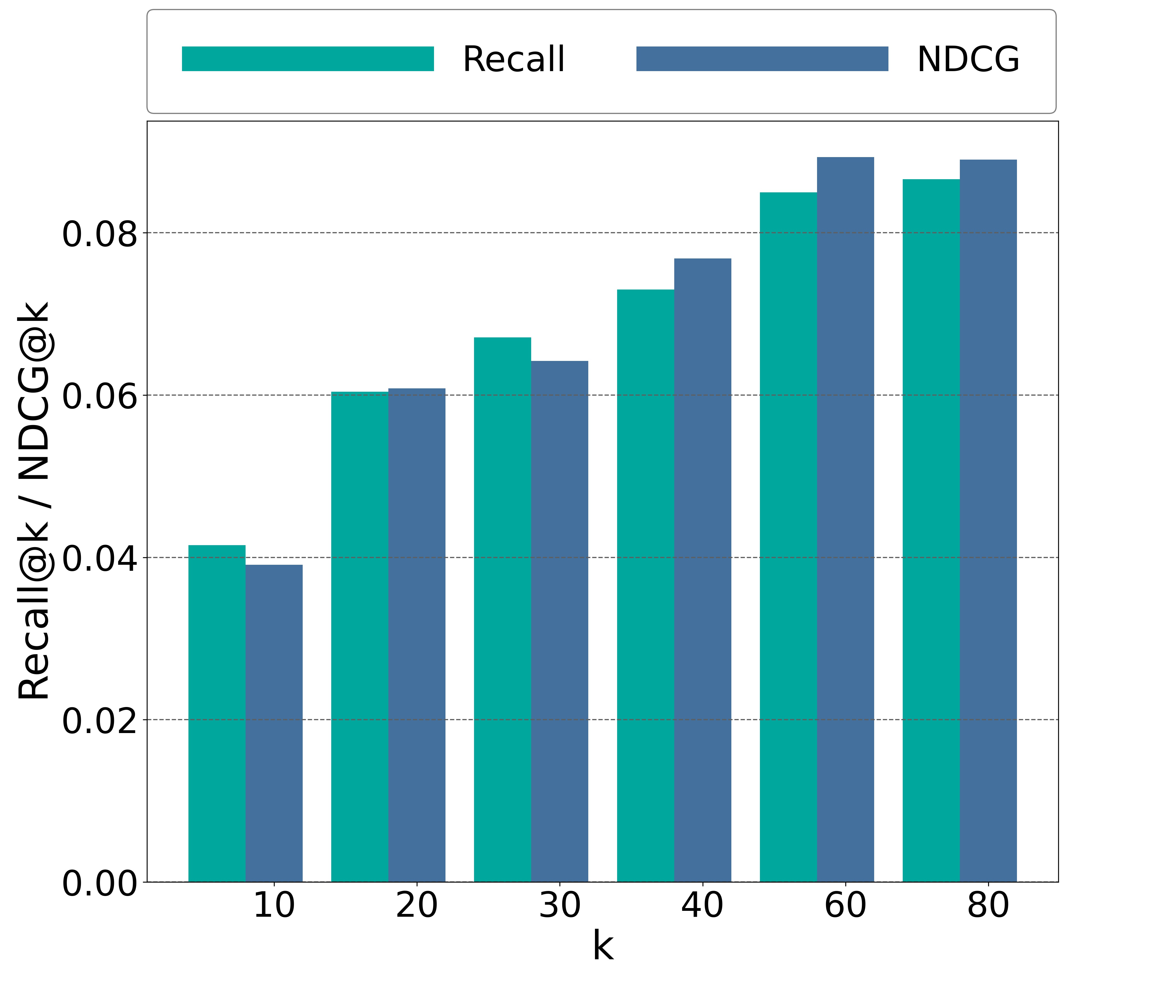}
\subcaption{(a) \texttt{MovieLens-HetRec}.}
\end{minipage}
\begin{minipage}[t]{0.15\textwidth}
\centering
\includegraphics[width=\textwidth]{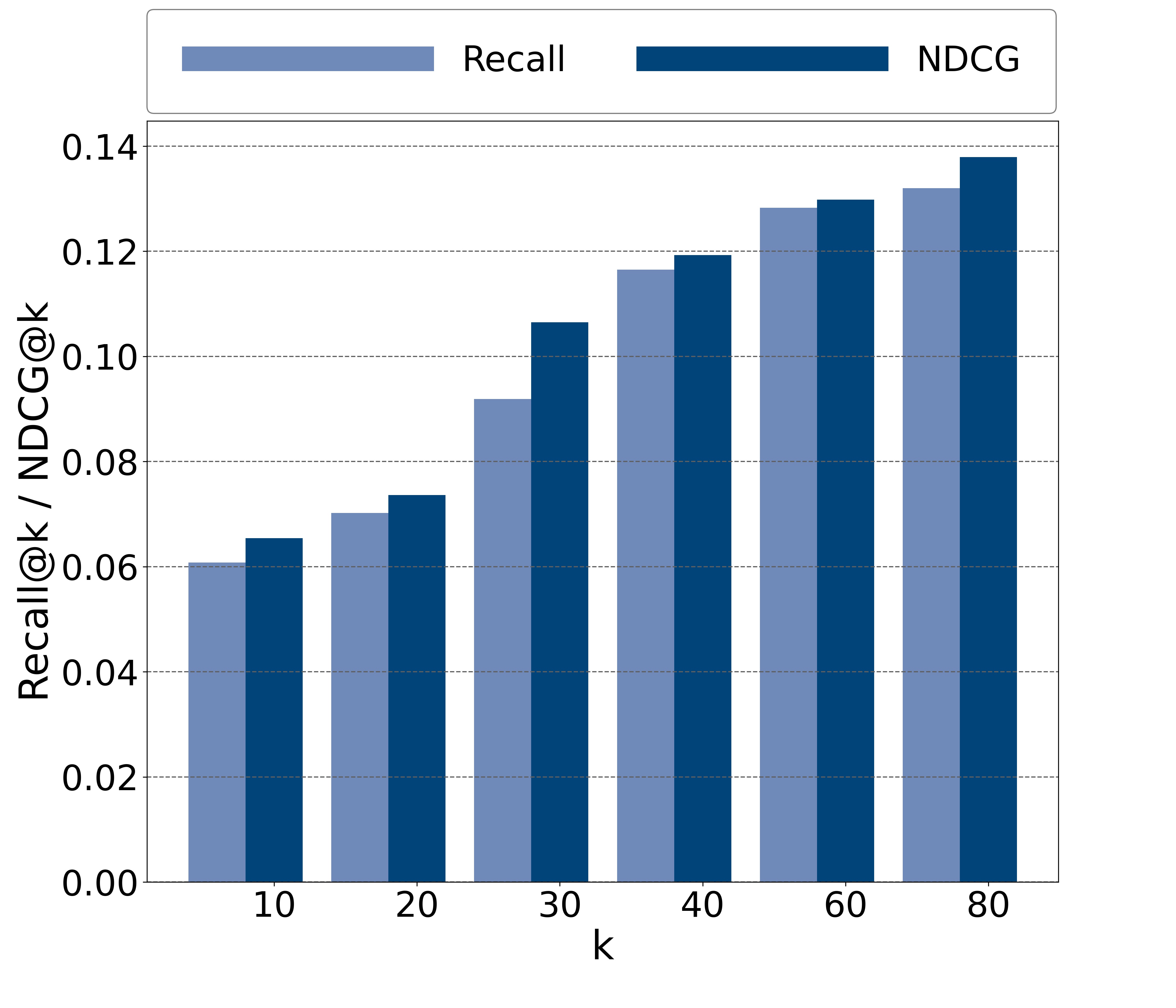}
\subcaption{(b) \texttt{Douban Book}.}
\end{minipage}
\begin{minipage}[t]{0.15\textwidth}
\centering
\includegraphics[width=\textwidth]{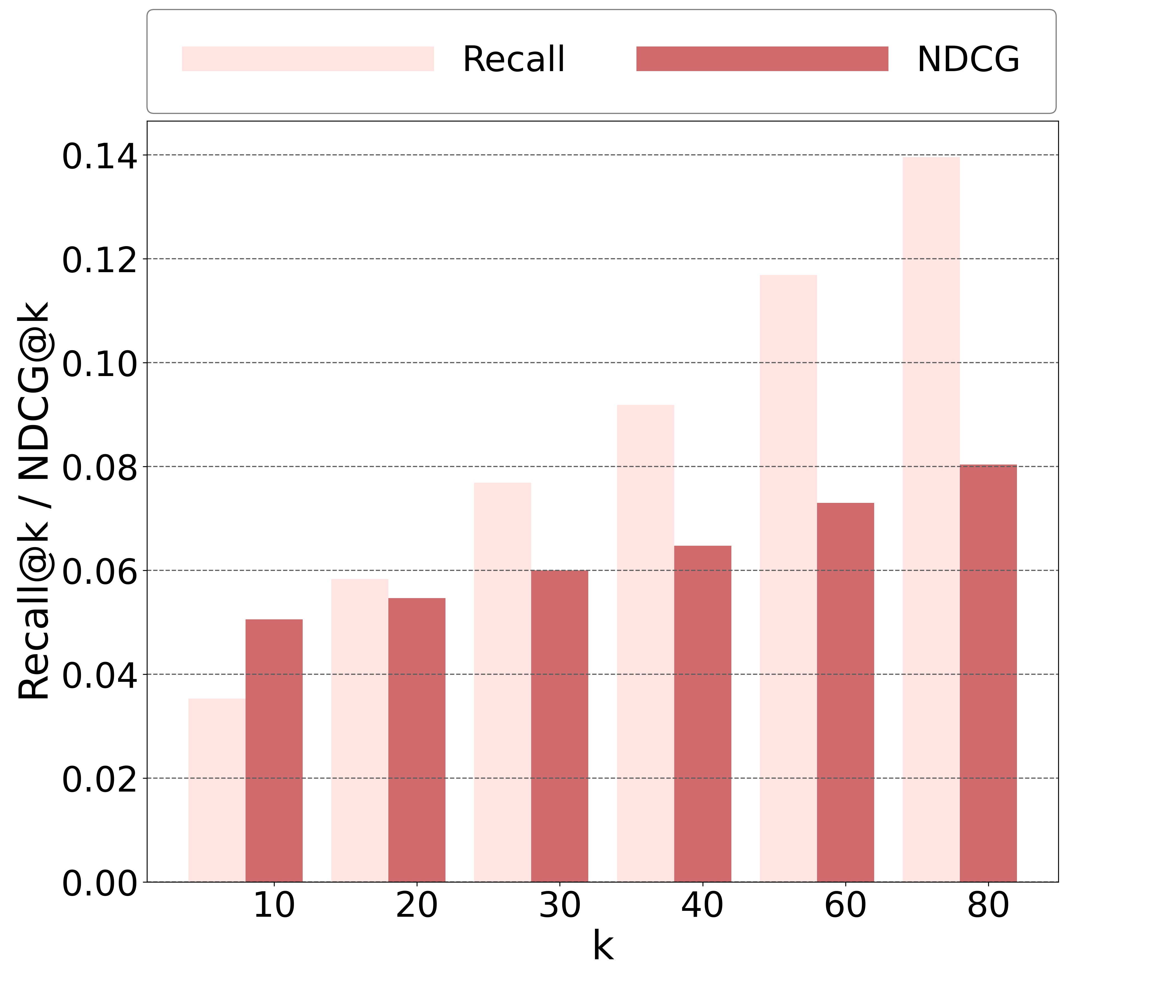}
\subcaption{(c) \texttt{Douban Movie}.}
\end{minipage}
\caption{Performance of CaDSI in terms of Recall@K and NDCG@K under different $K$.}
\label{fig:rec_k}
\end{figure}

Based on the evaluation on Recall@$K$ and NDCG@$K$, Figure~\ref{fig:rec_k} shows that CaDSI achieves the stable performance on top-$K$ recommendation when $K$ (i.e., the length of ranking list) varies from 10 to 80.
This indicates that our CaDSI performs stably on top-$K$ recommendation task and can recommend more relevant items within top-$K$ positions when the ranking list length increases. 

\subsection{Case Study and Visualization(RQ4)}
We conduct experiments to get deep insights into the disentangled representations w.r.t. the disentanglement of the semantics of user intents, the representability and interpretability of the learned disentangled embedding. 
The case studies towards the disentanglement of the semantics of user intents are shown in Appendix~\ref{app:rq4_1}, the 
visualization result of the learned disentangled embedding is shown in Appendix~\ref{app:rq4_2}.

\section{Related Work}
In this section, we will introduce previous works related to
ours from the following three aspects, including
HIN enhanced representation, disentangled representation and causal inference for recommendation.

\subsection{HIN Enhanced Representation}

As a newly emerging direction, heterogeneous information
network~\cite{shi2016survey} is proved to be effective in modeling complex objects and providing rich semantics information to recommender systems~\cite{shi2016survey,sun2013mining}.
Many HIN-based recommendation methods achieve the-state-of-the-art performance~\cite{shi2016survey}. 
For example,
HeteMF~\cite{yu2013collaborative} utilizes meta path based similarities as regularization terms in the MF model. 
HeteRec~\cite{yu2014personalized} learns meta path based latent features based on different types of entity relationships and proposes an enhanced personalized recommendation framework.
SemRec~\cite{shi2015semantic} proposes a weighted HIN and designs a meta path based CF model to flexibly integrate heterogeneous information for a personalized recommendation.
The effectiveness of HIN has been proved by a vast amount of HIN-based recommendation methods~\cite{shi2016survey}, thus, in our work, we value HIN in providing rich semantics information of user and item types.
Despite the effectiveness,  neither the enhanced graph-based nor HIN-based representations can disentangle users' intents by just presuming a uniform entangled embeddings behind behaviors.
This can result in the poor interpretability of the developed recommendation methods.
Thus, disentangle representation learning, which aims to learn factorized representations that separate and uncover latent explanatory factors behind the data~\cite{bengio2013representation}, has recently received much attention in recommendation systems. 

\subsection{Disentangled Representation}
Previous study has demonstrated that disentangled representations are more robust, i.e., counfounding bias are less likely to be preserved by uncovering latent factors.
Ma et al.~\cite{ma2019learning} propose to differentiate latent factors of learned user/item embeddings into macro and micro ones, thus the developed recommendation methods are less likely to mistakenly preserve the confounding of the factors.
Moreover, the disentangle representation can provide rich semantics of users' preference, involving items' aspect information~\cite{zhang2014explicit,ma2019learning} to users' behavior type information~\cite{liu2013adreveal}.
Thus, several works are proposed using disentangle representation learning to improve recommendation, 
for instance, DisHAN~\cite{wang2020disenhan} learns disentangled aspect-aware user/item representations based on different meta path types in a HIN, these aspect-aware embeddings are then used to guide the top-N recommendation.
Unfortunately, these aspect-aware disentangled embeddings only captured users' general taste on item aspects, however, failed to combine the specified item aspects with the real user intents.
Parallelly, several works are conducted on modeling disentangled representations of users' intents, such as MacridVAE~\cite{ma2019learning}, DICE~\cite{10.1145/3442381.3449788} and DGCF~\cite{wang2020disentangled}, the drawback is also distinct that they failed to combine the learned user intents with real-world item aspects, i.e., the user intents are predefined manually, such as "passing the time", short of providing meaningful information in a recommendation method. 
To sum up, the current studies on disentangle representation learning-enhanced recommendation either target at learning items' aspect-level representation~\cite{han2018aspect,wang2020disenhan} or users' intents-level representation~\cite{ma2019learning,10.1145/3442381.3449788,wang2020disentangled}.
However, aforementioned approaches ignore bias stemmed from semantics information. To our knowledge, our approach is the first attempt to achieve interpretable and unbiased recommendation with disentangled embeddings for user intent.

\subsection{Causal Methods for Debiasing}
To the best of our knowledge, existing causal methods for recommendations aim at mitigating the effects of different bias rather than improving interpretability as in our work.
Most existing works claim that the observational rating data suffers from \emph{selection bias}~\cite{yao2018representation,li2021causal}, 
\emph{exposure bias}~\cite{liang2016causal,schnabel2016recommendations,wang2018deconfounded,bonner2018causal} 
or 
\emph{popularity bias}~\cite{zheng2021disentangling,zhang2021causal,wang2021deconfounded}.
Following this paradigm, dominant approaches adopt two main strategies such as propensity-score~\cite{liang2016causal,schnabel2016recommendations,gruson2019offline} or causal embedding~\cite{wang2018deconfounded, yao2018representation,bonner2018causal}, to disentangle user interests from different types of bias.
For instance, the method in \cite{schnabel2016recommendations} uses propensity score to re-weight the observational click data, with the aim of imitating the scenario that item is randomly exposed and alleviating the \emph{exposure bias}. 
The work in ~\cite{wang2018deconfounded} learns a uniform unbiased embeddings from partially observed user-item interactions via their decounfonded model.
More recently, the work in~\cite{yao2018representation} resorts to balance learning with a Middle-point Distance Minimization (MPDM) strategy to learn causal embeddings that are free from \emph{selection bias}.
Facing user conformity issue in recommendation, ~\cite{zheng2021disentangling} relates such issue with \emph{popularity bias}, and proposes to alleviate the popularity bias by learning disentangled embeddings of user interest.
A few state-of-the-art works~\cite{wang2021deconfounded,li2021causal,zhang2021causal} inspect cause-effect of the bias generation and design a specific causal graph attributing the \emph{exposure bias} to a confounder. 
For example, Li et al.~\cite{li2021causal} prove the social network to be a confounder that affects the user's rating and the exposure policy of the item to the user.

\section{Conclusion and Future Work}
In this paper, we have researched the confounding bias issue stemming from different aspects, and propose an unbiased and robust \emph{Causal Disentanglement Semantics-Aware Intent Learning} (CaDSI) for recommendation.
Our CaDSI is capable of providing semantics to fine-grained representations for disentangling user intents, meanwhile easing the bias stemming from unevenly distributed item aspects.
We evaluate our CaDSI on three real-world recommendation datasets, 
with extensive experiments and visualizations demonstrate the robustness and interpretability of our semantics-aware user intent representation.
In future work, we will explore the effect of different auxiliary information on the recommendation system using the intervention analysis in causal inference.

\bibliographystyle{IEEEtran}
\bibliography{bibs}

\begin{thebibliography}{10}
\providecommand{\url}[1]{#1}
\csname url@samestyle\endcsname
\providecommand{\newblock}{\relax}
\providecommand{\bibinfo}[2]{#2}
\providecommand{\BIBentrySTDinterwordspacing}{\spaceskip=0pt\relax}
\providecommand{\BIBentryALTinterwordstretchfactor}{4}
\providecommand{\BIBentryALTinterwordspacing}{\spaceskip=\fontdimen2\font plus
\BIBentryALTinterwordstretchfactor\fontdimen3\font minus
  \fontdimen4\font\relax}
\providecommand{\BIBforeignlanguage}[2]{{%
\expandafter\ifx\csname l@#1\endcsname\relax
\typeout{** WARNING: IEEEtran.bst: No hyphenation pattern has been}%
\typeout{** loaded for the language `#1'. Using the pattern for}%
\typeout{** the default language instead.}%
\else
\language=\csname l@#1\endcsname
\fi
#2}}
\providecommand{\BIBdecl}{\relax}
\BIBdecl

\bibitem{koren2009matrix}
Y.~Koren, R.~Bell, and C.~Volinsky, ``Matrix factorization techniques for
  recommender systems,'' \emph{Computer}, vol.~42, no.~8, pp. 30--37, 2009.

\bibitem{salakhutdinov2007restricted}
R.~Salakhutdinov, A.~Mnih, and G.~Hinton, ``Restricted boltzmann machines for
  collaborative filtering,'' in \emph{Proceedings of the 24th international
  conference on Machine learning}, 2007, pp. 791--798.

\bibitem{zheng2021disentangling}
Y.~Zheng, C.~Gao, X.~Li, X.~He, Y.~Li, and D.~Jin, ``Disentangling user
  interest and conformity for recommendation with causal embedding,'' in
  \emph{Proceedings of the Web Conference 2021}, 2021, pp. 2980--2991.

\bibitem{wang2021deconfounded}
W.~Wang, F.~Feng, X.~He, X.~Wang, and T.-S. Chua, ``Deconfounded recommendation
  for alleviating bias amplification,'' \emph{arXiv preprint arXiv:2105.10648},
  2021.

\bibitem{gruson2019offline}
A.~Gruson, P.~Chandar, C.~Charbuillet, J.~McInerney, S.~Hansen, D.~Tardieu, and
  B.~Carterette, ``Offline evaluation to make decisions about
  playlistrecommendation algorithms,'' in \emph{Proceedings of the Twelfth ACM
  International Conference on Web Search and Data Mining}, 2019, pp. 420--428.

\bibitem{schnabel2016recommendations}
T.~Schnabel, A.~Swaminathan, A.~Singh, N.~Chandak, and T.~Joachims,
  ``Recommendations as treatments: Debiasing learning and evaluation,'' in
  \emph{international conference on machine learning}.\hskip 1em plus 0.5em
  minus 0.4em\relax PMLR, 2016, pp. 1670--1679.

\bibitem{wang2018deconfounded}
Y.~Wang, D.~Liang, L.~Charlin, and D.~Blei, ``The deconfounded recommender: A
  causal inference approach to recommendation,'' 08 2018.

\bibitem{ma2019learning}
J.~Ma, C.~Zhou, P.~Cui, H.~Yang, and W.~Zhu, ``Learning disentangled
  representations for recommendation,'' \emph{NeurIPS}, 2019.

\bibitem{li2021causal}
Q.~Li, X.~Wang, and G.~Xu, ``Be causal: De-biasing social network confounding
  in recommendation,'' \emph{arXiv preprint arXiv:2105.07775}, 2021.

\bibitem{zhang2021causal}
Y.~Zhang, F.~Feng, X.~He, T.~Wei, C.~Song, G.~Ling, and Y.~Zhang, ``Causal
  intervention for leveraging popularity bias in recommendation,'' \emph{arXiv
  preprint arXiv:2105.06067}, 2021.

\bibitem{10.1145/3442381.3449788}
\BIBentryALTinterwordspacing
Y.~Zheng, C.~Gao, X.~Li, X.~He, Y.~Li, and D.~Jin, ``Disentangling user
  interest and conformity for recommendation with causal embedding,'' in
  \emph{Proceedings of the Web Conference 2021}, ser. WWW '21.\hskip 1em plus
  0.5em minus 0.4em\relax New York, NY, USA: Association for Computing
  Machinery, 2021, p. 2980–2991. [Online]. Available:
  \url{https://doi.org/10.1145/3442381.3449788}
\BIBentrySTDinterwordspacing

\bibitem{ying2018graph}
R.~Ying, R.~He, K.~Chen, P.~Eksombatchai, W.~L. Hamilton, and J.~Leskovec,
  ``Graph convolutional neural networks for web-scale recommender systems,'' in
  \emph{Proceedings of the 24th ACM SIGKDD International Conference on
  Knowledge Discovery \& Data Mining}, 2018, pp. 974--983.

\bibitem{he2020lightgcn}
X.~He, K.~Deng, X.~Wang, Y.~Li, Y.~Zhang, and M.~Wang, ``Lightgcn: Simplifying
  and powering graph convolution network for recommendation,'' in
  \emph{Proceedings of the 43rd International ACM SIGIR Conference on Research
  and Development in Information Retrieval}, 2020, pp. 639--648.

\bibitem{shi2016survey}
C.~Shi, Y.~Li, J.~Zhang, Y.~Sun, and S.~Y. Philip, ``A survey of heterogeneous
  information network analysis,'' \emph{IEEE Transactions on Knowledge and Data
  Engineering}, vol.~29, no.~1, pp. 17--37, 2016.

\bibitem{pearl2009causality}
J.~Pearl, \emph{Causality}.\hskip 1em plus 0.5em minus 0.4em\relax Cambridge
  university press, 2009.

\bibitem{dong2017metapath2vec}
Y.~Dong, N.~V. Chawla, and A.~Swami, ``metapath2vec: Scalable representation
  learning for heterogeneous networks,'' in \emph{Proceedings of the 23rd ACM
  SIGKDD international conference on knowledge discovery and data mining},
  2017, pp. 135--144.

\bibitem{grover2016node2vec}
A.~Grover and J.~Leskovec, ``node2vec: Scalable feature learning for
  networks,'' in \emph{Proceedings of the 22nd ACM SIGKDD international
  conference on Knowledge discovery and data mining}, 2016, pp. 855--864.

\bibitem{bottou2012stochastic}
L.~Bottou, ``Stochastic gradient descent tricks,'' in \emph{Neural networks:
  Tricks of the trade}.\hskip 1em plus 0.5em minus 0.4em\relax Springer, 2012,
  pp. 421--436.

\bibitem{weisberg2005applied}
S.~Weisberg, \emph{Applied linear regression}.\hskip 1em plus 0.5em minus
  0.4em\relax John Wiley \& Sons, 2005, vol. 528.

\bibitem{berg2017graph}
R.~v.~d. Berg, T.~N. Kipf, and M.~Welling, ``Graph convolutional matrix
  completion,'' \emph{KDD}, 2017.

\bibitem{wang2019neural}
X.~Wang, X.~He, M.~Wang, F.~Feng, and T.-S. Chua, ``Neural graph collaborative
  filtering,'' in \emph{Proceedings of the 42nd international ACM SIGIR
  conference on Research and development in Information Retrieval}, 2019, pp.
  165--174.

\bibitem{sharma2017activation}
S.~Sharma and S.~Sharma, ``Activation functions in neural networks,''
  \emph{Towards Data Science}, vol.~6, no.~12, pp. 310--316, 2017.

\bibitem{5694074}
S.~Rendle, ``Factorization machines,'' in \emph{2010 IEEE International
  Conference on Data Mining}, 2010, pp. 995--1000.

\bibitem{zhang2019stylistic}
S.~Zhang, Z.~Han, Y.-K. Lai, M.~Zwicker, and H.~Zhang, ``Stylistic scene
  enhancement gan: mixed stylistic enhancement generation for 3d indoor
  scenes,'' \emph{The Visual Computer}, vol.~35, no.~6, pp. 1157--1169, 2019.

\bibitem{lian2020personalized}
D.~Lian, Q.~Liu, and E.~Chen, ``Personalized ranking with importance
  sampling,'' in \emph{Proceedings of The Web Conference 2020}, 2020, pp.
  1093--1103.

\bibitem{bushaev2018adam}
V.~Bushaev, ``Adam—latest trends in deep learning optimization,''
  \emph{Towards Data Science, Listopad}, 2018.

\bibitem{he2017neural}
X.~He, L.~Liao, H.~Zhang, L.~Nie, X.~Hu, and T.-S. Chua, ``Neural collaborative
  filtering,'' in \emph{Proceedings of the 26th international conference on
  world wide web}, 2017, pp. 173--182.

\bibitem{yu2018adaptive}
J.~Yu, M.~Gao, J.~Li, H.~Yin, and H.~Liu, ``Adaptive implicit friends
  identification over heterogeneous network for social recommendation,'' in
  \emph{Proceedings of the 27th ACM international conference on information and
  knowledge management}, 2018, pp. 357--366.

\bibitem{hu2018leveraging}
B.~Hu, C.~Shi, W.~X. Zhao, and P.~S. Yu, ``Leveraging meta-path based context
  for top-n recommendation with a neural co-attention model,'' in
  \emph{Proceedings of the 24th ACM SIGKDD International Conference on
  Knowledge Discovery \& Data Mining}, 2018, pp. 1531--1540.

\bibitem{han2018aspect}
X.~Han, C.~Shi, S.~Wang, S.~Y. Philip, and L.~Song, ``Aspect-level deep
  collaborative filtering via heterogeneous information networks.'' in
  \emph{IJCAI}, 2018, pp. 3393--3399.

\bibitem{wang2020disentangled}
X.~Wang, H.~Jin, A.~Zhang, X.~He, T.~Xu, and T.-S. Chua, ``Disentangled graph
  collaborative filtering,'' in \emph{Proceedings of the 43rd International ACM
  SIGIR Conference on Research and Development in Information Retrieval}, 2020,
  pp. 1001--1010.

\bibitem{glorot2010understanding}
X.~Glorot and Y.~Bengio, ``Understanding the difficulty of training deep
  feedforward neural networks,'' in \emph{Proceedings of the thirteenth
  international conference on artificial intelligence and statistics}, 2010,
  pp. 249--256.

\bibitem{sun2013mining}
Y.~Sun and J.~Han, ``Mining heterogeneous information networks: a structural
  analysis approach,'' \emph{Acm Sigkdd Explorations Newsletter}, vol.~14,
  no.~2, pp. 20--28, 2013.

\bibitem{yu2013collaborative}
X.~Yu, X.~Ren, Q.~Gu, Y.~Sun, and J.~Han, ``Collaborative filtering with entity
  similarity regularization in heterogeneous information networks,''
  \emph{IJCAI HINA}, vol.~27, 2013.

\bibitem{yu2014personalized}
X.~Yu, X.~Ren, Y.~Sun, Q.~Gu, B.~Sturt, U.~Khandelwal, B.~Norick, and J.~Han,
  ``Personalized entity recommendation: A heterogeneous information network
  approach,'' in \emph{Proceedings of the 7th ACM international conference on
  Web search and data mining}, 2014, pp. 283--292.

\bibitem{shi2015semantic}
C.~Shi, Z.~Zhang, P.~Luo, P.~S. Yu, Y.~Yue, and B.~Wu, ``Semantic path based
  personalized recommendation on weighted heterogeneous information networks,''
  in \emph{Proceedings of the 24th ACM International on Conference on
  Information and Knowledge Management}, 2015, pp. 453--462.

\bibitem{bengio2013representation}
Y.~Bengio, A.~Courville, and P.~Vincent, ``Representation learning: A review
  and new perspectives,'' \emph{IEEE transactions on pattern analysis and
  machine intelligence}, vol.~35, no.~8, pp. 1798--1828, 2013.

\bibitem{zhang2014explicit}
Y.~Zhang, G.~Lai, M.~Zhang, Y.~Zhang, Y.~Liu, and S.~Ma, ``Explicit factor
  models for explainable recommendation based on phrase-level sentiment
  analysis,'' in \emph{Proceedings of the 37th international ACM SIGIR
  conference on Research \& development in information retrieval}, 2014, pp.
  83--92.

\bibitem{liu2013adreveal}
B.~Liu, A.~Sheth, U.~Weinsberg, J.~Chandrashekar, and R.~Govindan, ``Adreveal:
  Improving transparency into online targeted advertising,'' in
  \emph{Proceedings of the Twelfth ACM Workshop on Hot Topics in Networks},
  2013, pp. 1--7.

\bibitem{wang2020disenhan}
Y.~Wang, S.~Tang, Y.~Lei, W.~Song, S.~Wang, and M.~Zhang, ``Disenhan:
  Disentangled heterogeneous graph attention network for recommendation,'' in
  \emph{Proceedings of the 29th ACM International Conference on Information \&
  Knowledge Management}, 2020, pp. 1605--1614.

\bibitem{yao2018representation}
L.~Yao, S.~Li, Y.~Li, M.~Huai, J.~Gao, and A.~Zhang, ``Representation learning
  for treatment effect estimation from observational data,'' \emph{Advances in
  Neural Information Processing Systems}, vol.~31, 2018.

\bibitem{liang2016causal}
D.~Liang, L.~Charlin, and D.~M. Blei, ``Causal inference for recommendation,''
  in \emph{Causation: Foundation to Application, Workshop at UAI}.\hskip 1em
  plus 0.5em minus 0.4em\relax AUAI, 2016.

\bibitem{bonner2018causal}
S.~Bonner and F.~Vasile, ``Causal embeddings for recommendation,'' in
  \emph{Proceedings of the 12th ACM Conference on Recommender Systems}, 2018,
  pp. 104--112.

\bibitem{van2008visualizing}
L.~Van~der Maaten and G.~Hinton, ``Visualizing data using t-sne.''
  \emph{Journal of machine learning research}, vol.~9, no.~11, 2008.

\end{thebibliography}
\begin{IEEEbiography}[{\includegraphics[width=1in,height=1.25in,clip,keepaspectratio]{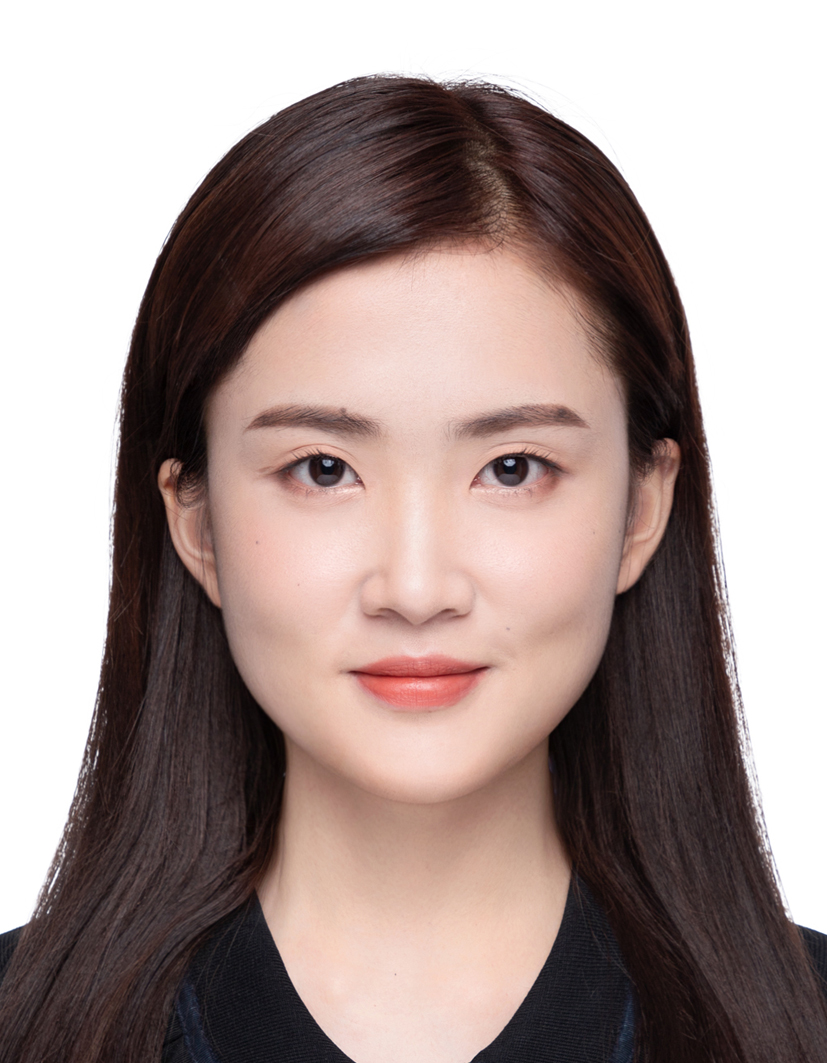}}]{Xiangmeng Wang} has been a Ph.D. student at the School of Computer Science, Faculty of Engineering and Information Technology, University of Technology Sydney (UTS). She received her MSc degree in Computer Application Technology from Shanghai University. Her general research interests lie primarily in explainable artificial intelligence, data analysis, and causal machine learning. Her papers have been published in the top-tier conferences and journals in the field of machine learning.
\end{IEEEbiography}

\begin{IEEEbiography}[{\includegraphics[width=1in,height=1.25in,clip,keepaspectratio]{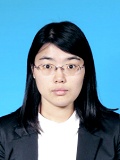}}]{Qian Li} is a Lecturer at School of Electrical Engineering, Computing and Mathematical
Sciences, Curtin University, Perth, Australia.
She has been a Postdoc Research Fellow at the School of Computer Science, Faculty of Engineering and Information Technology, University of Technology Sydney (UTS). She received her Ph.D. in Computer Science from the Chinese Academy of Science. Her general research interests lie primarily in optimization algorithms, topological data analysis, and causal machine learning. Her papers have been published in the top-tier conferences and journals in the field of machine learning and computer vision.
\end{IEEEbiography}

\begin{IEEEbiography}[{\includegraphics[width=1in,height=1.25in,clip,keepaspectratio]{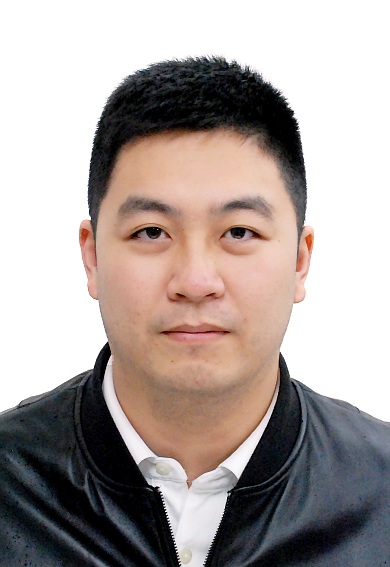}}]{Dianer Yu} has been a Postgraduate IT student at the School of Computer Science, Faculty of Engineering and Information Technology, University of Technology Sydney (UTS). He received his BSc degree of IT from University of Technology Sydney. His general research interests lie primarily in data mining, causal model for recommendation and explainable machine learning. He has been awarded as Postgraduate Dean's List during the Postgraduate period.
\end{IEEEbiography}

\begin{IEEEbiography}[{\includegraphics[width=1in,height=1.25in,clip,keepaspectratio]{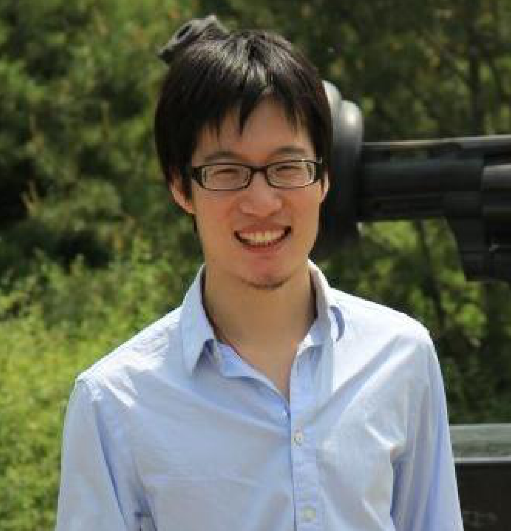}}]{Peng Cui} is an Associate Professor at Tsinghua University. He received his Ph.D. degree in computer science in 2010 from Tsinghua University. He has vast research interests in data mining, multimedia processing, and social network analysis. Until now, he has published more than 20 papers in conferences such as SIGIR, AAAI, ICDM, etc. and journals such as IEEE TMM, IEEE TIP, DMKD, etc. Now his research is sponsored by National Science Foundation of China, Samsung, Tencent, etc. He also serves as Guest Editor, Co-Chair, PC member, and Reviewer of several high-level international conferences, workshops, and journals.
\end{IEEEbiography}

\begin{IEEEbiography}[{\includegraphics[width=1in,height=1.25in,clip,keepaspectratio]{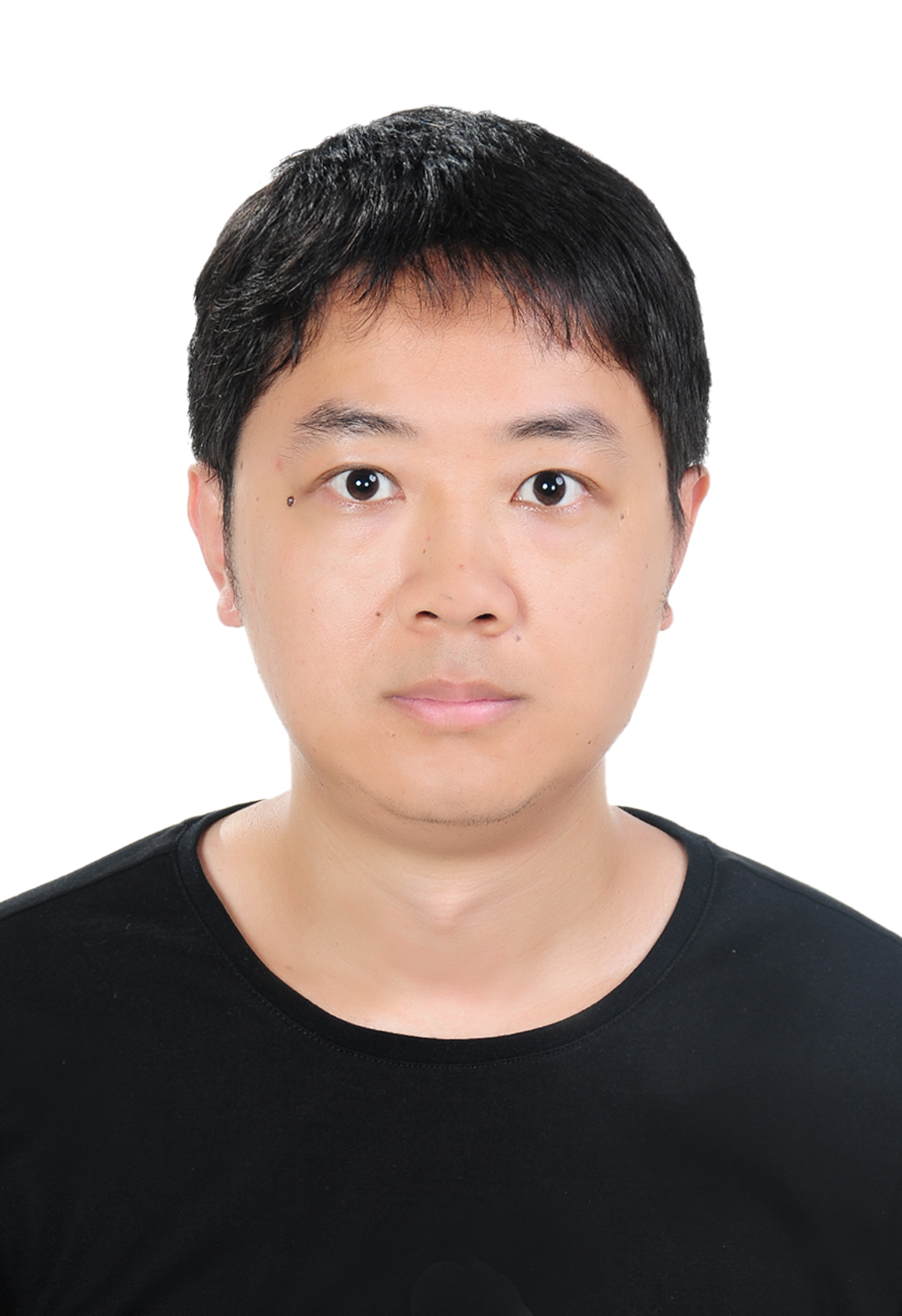}}]
{Zhichao Wang} received his Ph.D. degree from Department of Automation, Tsinghua University. He was a Research Fellow at University of New South Wales. His research interests lie in the optimization for machine learning and stochastic modeling. 
\end{IEEEbiography}

\begin{IEEEbiography}[{\includegraphics[width=1in,height=1.25in,clip,keepaspectratio]{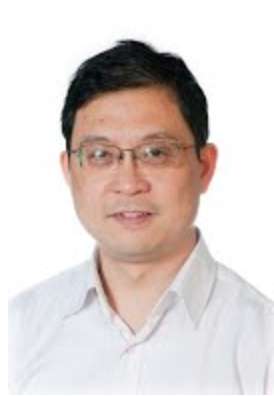}}]{Guandong Xu} is a Professor in the School of Computer Science and Advanced Analytics Institute at University of Technology Sydney. He received MSc and BSc degree in Computer Science and Engineering, and PhD in Computer Science. He currently heads the Data Science and Machine Intelligence Lab, which consists of 15+ members of academics, research fellows and HDR students. From Nov 2019, he directs the newly established Smart Future Research Centre, which is an across-disciplines industry engagement and innovation platform for AI and Data Science Application towards smart wealth management and investment, energy, food, water, living, and city.
\end{IEEEbiography}

\newpage
\appendices

\section{Case Studies}\label{app:rq4_1}
\begin{figure}[htbp]
\centering
    \begin{minipage}[t]{0.45\textwidth}
        \centering
        \includegraphics[width=\textwidth]{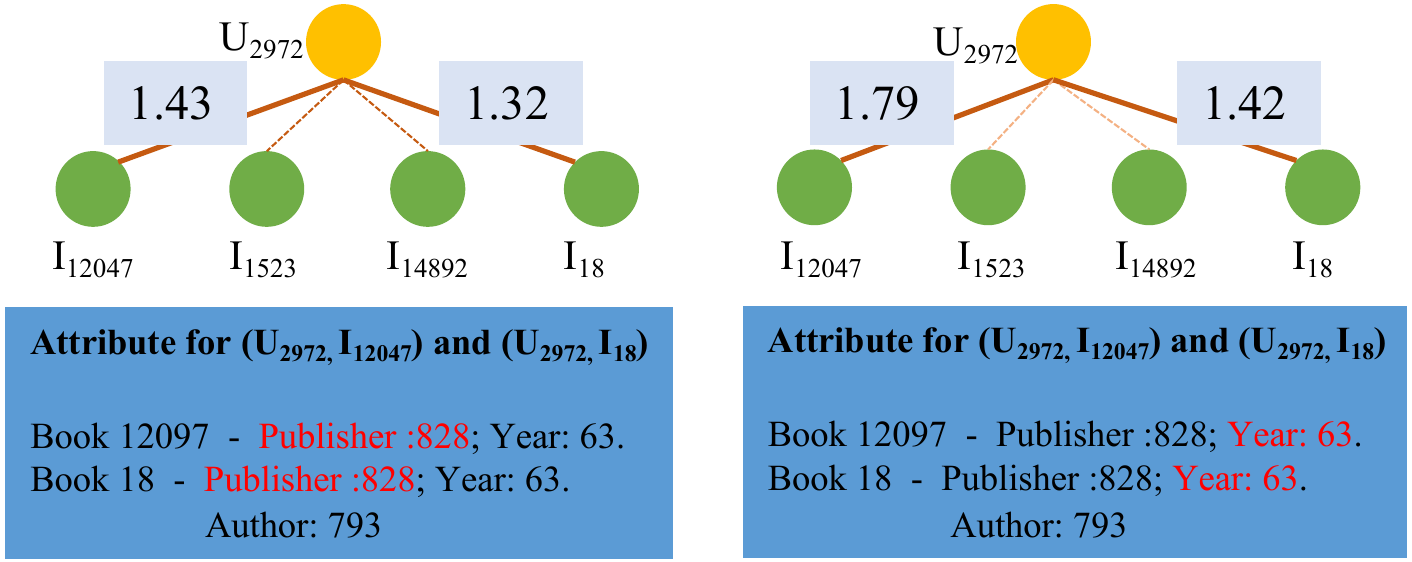}
    \end{minipage}
        \begin{minipage}[t]{0.45\textwidth}
        \centering
      \includegraphics[width=\textwidth]{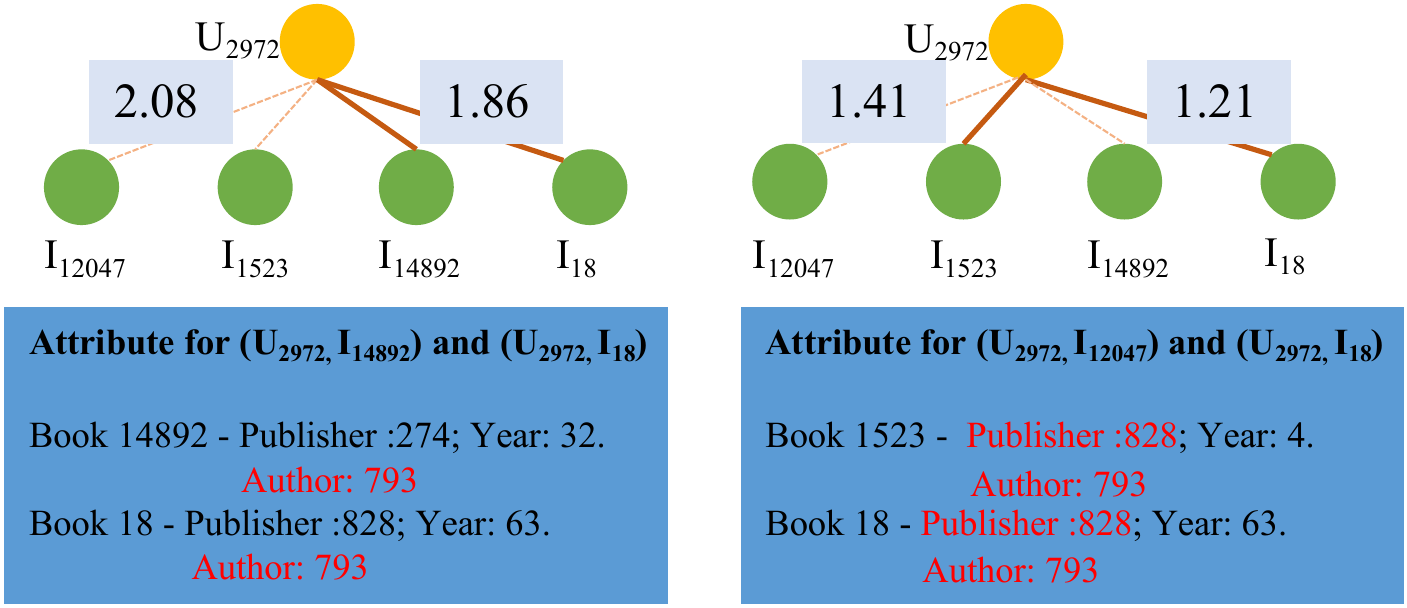}
    \end{minipage}
    \caption{Visualization of the disentangled user intent graphs based on score matrices. user-item interactions with highest scores are marked in solidlines; item attributes with the same values are highlighted in red.}
    \label{fig:case_1}
\end{figure}

We first conduct an experiment to understand the disentanglement of user intents by our CaDSI, then explore whether such intent related to real-world item semantics.
We select an user $u2972$ from \texttt{Douban Book} and learn its interaction scores $\mathbf{S}(u, i)$ (\textit{cf.} Eq.~\eqref{one_layer}) with his/her historical interacted items under our CaDSI. 
The user intents factor $k=4$  indicates four distinct user intents. 
Thereafter, we randomly select four items from the interaction score matrices. 
For each interaction under different $k$, we mark the interaction scores with the highest confidence with solid lines and couple the certain item attributes below them. 
Figure~\ref{fig:case_1} shows the visualization results and we have the following findings:

\begin{itemize}
    \item Jointly analyzing intent-aware user-item interaction graphs, we can see
    user preference differs across each graphs, reflected by different interaction scores in each intent-aware graph. 
    For example, $u2972$ interacts with $i12047$ with a preference score of $1.43$ under intent $k_1$, while the score changes to $1.79$ under intent $k_2$.
    This demonstrates the importance of disentangling user intents in recommendation scenario. 
    \item We thereafter couple item attributes to investigate whether user intents are related to item semantics. 
    It can be seen that different fine-grained user intents are highly consistent with high-level item semantics.
    For instance, intent $k_3$ contributes mostly to interactions $(u2972,i14892)$ and $(u2972,i18)$, which suggests its high confidence as being the intents behind these behaviors. When switch to item attributes of $i12047$ and $i18$, one highlighted item attribute \emph{Author} with the same id can be found, reflecting the reason why $u2972$ chose to interact with $i12047$ and $i18$.
    This demonstrated that our CaDSI, which aims at disentangling user intents meanwhile assign specific item semantics to the learned intents, is effective in the disentanglement of user intents towards item aspects.
\end{itemize}

\section{Visualization}\label{app:rq4_2}

\begin{figure}[htbp]
\centering
    \begin{minipage}[t]{0.5\textwidth}
        \centering
        \includegraphics[width=0.5\textwidth]{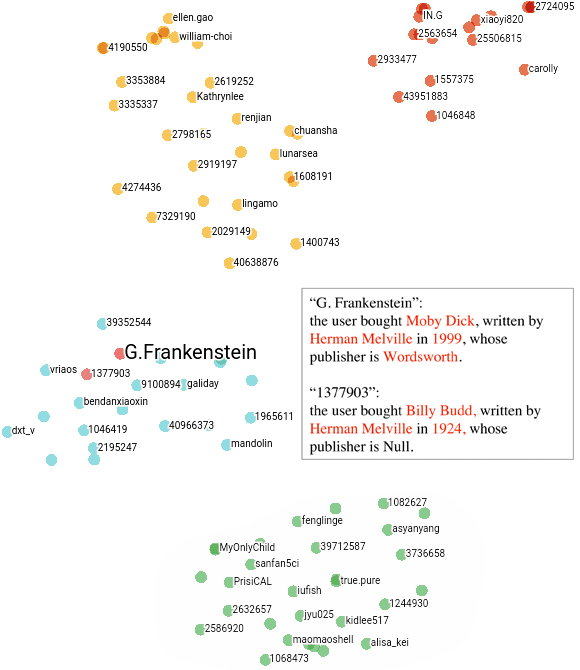}
    \end{minipage}
    \caption{$2$-dimensional t-SNE projections of the $128$-dimensional embeddings of $100$ users from \texttt{Douban Book} dataset.}
    \label{fig:case_2}
\end{figure}

We randomly select $100$ users from \texttt{Douban Book} dataset, and implement CaDSI on the dataset to output the $128$-dimensional semantics-aware user intent embedding $\boldsymbol{e}$.
For visualization purposes, we use \texttt{t-SNE}~\cite{van2008visualizing} to map high-dimensional user intent representation $\boldsymbol{e}$ to $2$-dimensional vectors.   
Following the parameter settings in Section~\ref{para}, 
we set the user intent factors $k=4$. 
Figure~\ref{fig:case_2} shows the visualization result.

We notice that the projections are capable of distinguishing four discernible clusters of users, and the cluster number is consistent with our pre-defined latent user intent factors $k$.
This indicates that CaDSI is able to group users of the same intent closely based on the distances among users' embeddings. 
Meanwhile, each cluster is well-separated from others, further demonstrating
the robust representation of CaDSI.

Furthermore, we extract two users termed $1377903$ and $G. Frankenstein$ and show their historical interaction abstracts in the right of Figure~\ref{fig:case_2}.
Analyzing these abstracts, we can see that our CaDSI is also capable of grouping each user whose interests are on the same item attributes. Such as $1377903$ and $G. Frankenstein$, both users arrange items with similar attributes close to each other and dissimilar ones distant from each other.
In summary, the visualizations intuitively demonstrate CaDSI’s novel capability to discover, model, and capture the underlying semantics and structural relationships between multiple item aspects and user intents in heterogeneous networks.

\end{document}